\documentclass[12pt]{article}
\usepackage{jheppub}
\usepackage{amsfonts,bbold}   
\usepackage{amssymb,amsmath}
\usepackage{cancel}
\usepackage{float}
\usepackage{xcolor}

\newcommand{\vt}{\vec \theta \, }
\newcommand{\vj}{\vec J}
\newcommand{\be}{\begin{equation}}
\newcommand{\ee}{\end{equation}}
\newcommand{\sfrac}[2]{\textstyle{\frac{#1}{#2}}}

\title{Rotating cosmologies: classical and quantum
}
\author[a]{Alberto Nicolis,}\emailAdd{a.nicolis@columbia.edu}
\author[b]{Federico Piazza,}\emailAdd{Federico.Piazza@cpt.univ-mrs.fr}
\author[b]{and Kenza Zeghari }\emailAdd{kenza.zeghari@univ-amu.fr}

\affiliation[a]{Center for Theoretical Physics and Department of Physics,\\
Columbia University, New York, NY 10027, USA}
\affiliation[b]{Aix Marseille Univ, Universit\'{e} de Toulon, CNRS, CPT, Marseille, France}

\abstract{We revisit spatially flat, anisotropic cosmologies within the framework of mini-superspace. Putting  special emphasis on the symmetries of the mini-superspace action and on the associated conservation laws, we unveil a new class of {\em rotating} cosmologies driven by solid matter. Their rotating is physical, in that it is characterized in an invariant way in terms of a conserved angular momentum. Along the way, we confirm the results of Bartolo et al.~regarding the slow decay of anisotropies for solid inflation.  We then use our minisuperspace approach as a laboratory to address certain puzzles of quantum cosmology---among these, how to characterize the spacetime symmetries of a quantum state at the level of the wavefunction of the universe. For the case of a solid driven cosmology, this question seems better defined than in more standard cases. Other questions remain unanswered, though; in particular, the general question of how to operate a minisuperspace-like truncation of degrees of freedom that is consistent at the quantum level.
}

\usepackage{hyperref}  
\hypersetup{
    colorlinks=true,
    linkcolor=blue,
    filecolor=magenta,      
    urlcolor=cyan,
}
\begin{document}
\maketitle
\numberwithin{equation}{section}  
\setcounter{tocdepth}{3}  
\newpage
\thispagestyle{empty}

\section{Introduction}

Primordial inflation is extremely efficient at turning generic initial conditions into a highly homogeneous and isotropic spacetime~\cite{Linde:1990flp}. Inhomogeneities are stretched out to unobservably large scales.  Random initial anisotropies are also rapidly diluted away by the quasi-exponential expansion. In this paper we point to another potential feature---``\emph{rotation}''---that inflation can remove and that, to our knowledge, has never been previously considered. 

In the homogeneous limit, anisotropies can be understood geometrically as a direction-dependent expansion rate.  Given a general spatially flat homogeneous metric
\begin{equation} \label{intro-1}
ds^2 = -dt^2 + h_{ij}(t) dx^i dx^j\, ,
\end{equation}
it is customary to choose coordinates such that $h_{ij}$ is diagonal, so that the different expansion rates can be easily read off. However, one may wonder, does $h_{ij}$ need to stay diagonal at all times? 
In the absence of anisotropic stress in the matter sector, it is easy to see that it does.

To this end, consider the spatial components of the Einstein equations for a metric in the form~\eqref{intro-1} in $d+1$ dimensions, 
\begin{equation}\label{eom0}
\ddot h_{ij} + \frac12 \dot h_{k l} h^{kl} \dot h_{ij} - \dot h_{i k} h^{kl} \dot h_{l j} \ = \ 16 \pi G\left(T_{ij} -\frac{1}{d- 1} \ h_{ij} T^\mu {}_\mu \right)\, .
\end{equation}  
The metric maintains the form~\eqref{intro-1} under any general coordinate-independent linear transformation. The invariance of~\eqref{eom0}
under $GL(d, \mathbb{R})$ is clearly a remnant of the original diffeomorphism invariance. By using an element of such a group, at some given instant $t = t_0$ the metric can be made proportional to the identity. At the same time, a suitable rotation can diagonalize $\dot h_{ij}(t_0)$ without affecting  $h_{ij}(t_0)$. From~\eqref{eom0} it is then clear that the metric will stay diagonal at all times, provided $T_{ij}$ does not contain off-diagonal entries. 

In the presence of anisotropic stress in the matter fields, however, the spatial metric does not remain diagonal at all times, and this corresponds to a non trivial ``rotation".  Rotation of what with respect to what? There is more than one way to answer this question but the most physical approach is probably to focus on the \emph{derivative}\footnote{A mathematically more convenient choice is to decompose the metric itself, as we do in the rest of the paper.} of the metric, $\dot h_{ij}$, which can be decomposed as
\begin{equation}
\dot h_{ij}(t) = \sum_{n =1}^d H_n(t) \  \hat u^{(n)}_i(t)\  \hat u^{(n)}_j (t)\, .
\end{equation} 
The eigenvalues $H_n$ are direction-dependent ``Hubble rates", while the instantaneous eigenvectors can be seen as the principal expansion axes, or as the principal axes of the \emph{extrinsic curvature} $K_{ij} \propto \dot h_{ij}$ of the hypersurfaces of constant $t$.
Because $\dot h_{ij}$ is symmetric, $\hat u^{(n)}_i(t)$ is, at any time, an orthonormal basis with respect to the standard Kronecker scalar product, 
\begin{equation}
\hat u^{(n)}_i(t) \, \hat u^{(m)}_j(t) \, \delta_{ij} = \delta_{m n}\, .
\end{equation}
So, ``rotation" means time-dependence of this orthogonal set of  principal axes FLRW
with respect to the system of comoving observers $x^i = {\rm const.}$~who are in geodesic motion. If one tried to define new spatial coordinates ${x' {}^i}$ in order to ``follow the rotation", such coordinates would not label geodesic observers any longer. 

In the rest of the paper (Secs.~\ref{rules} and~\ref{minisuperspace}), we revisit the usual mini-superspace approach sketched above by giving particular emphasis  to symmetries and the associated conservation laws. Such a viewpoint is useful because it allows us to recast the above analysis in terms of which charges can be set to zero by  a symmetry transformation. {\em If} the matter sector breaks certain symmetries of the gravitational action---as will be the case for a solid-driven cosmology---the angular momentum charges cannot be transformed to zero by a symmetry transformation. 
This, we take as the definition of having a ``rotating cosmology".
In Sec.~\ref{understand} we try to understand in physical terms the associated rotation, by specializing to $2+1$ dimensions, and we display some explicit  numerical solutions. We consider the quantum theory in Sec.~\ref{quantum2}.


\section{Classical minisuperspace actions: the rules of the game}\label{rules}

To study Friedmann Lema\^itre Robertson Walker (FLRW) cosmologies, one usually specializes the Einstein equations to the FLRW metric and to time-dependent matter fields, say in comoving coordinates and cosmic time, and tries to solve them.  An alternative approach, is to write the action in ``minisuperspace": that is, one substitutes the FLRW ansatz directly into the action, and interprets the resulting action as a functional of $a(t)$, of time-dependent matter fields, and of---famously---the lapse function $N(t)$. Indeed, retaining $N(t)$ as a degree of freedom is crucial to retain the Hamiltonian constraint as an equation of motion, which is nothing but the first Friedmann equation.

We would like to do the same for spatially flat, homogeneous, but anisotropic universes. Before attempting to do so, we must understand what the rules of the game are, in terms of how general an ansatz we should use, and in particular of how specializing to a given ansatz  interferes with gauge invariance: why do we have to keep $N(t)$? Should we also keep the shifts $N^i(t)$ in general?

Neglecting for a moment the subtleties associated with gauge invariance and gauge fixing, we recall a general argument of Coleman's for classical field theories: if an action is invariant under a symmetry group $G$, and one is looking for solutions to the field equations that preserve a subgroup $H \subseteq G$ of such symmetries, one can plug into the action the most general ansatz that is invariant under $H$, and vary the action within that subspace of field configurations. The two variational problems (first vary and then impose the symmetries of the solution, first impose the symmetries of the solution and then vary) yield the same set of solutions \cite{Coleman}.

How does one adapt the above argument to gauge theories, and in particular to our case? The answer is simple. For definiteness, let's consider directly the case of cosmological solutions in GR, which is what we are interested in. The symmetries of the action, $G$, are all the diffeomorphisms. The symmetries we would like our solution to preserve, $H \subset G$, are certain isometries, generated by some Killing vectors $\xi_H^\mu(x)$. However, in different gauges the Killing vectors take different forms. From the functional analysis viewpoint, which is at the basis of Coleman's argument, different-looking Killing vectors generate different symmetries. So, if we ask what is the most general field configuration that preserves a certain isometry $\xi_H^\mu(x)$, we must ask this with enough gauge-fixing to completely specify the functional form of $\xi_H^\mu(x)$.

For example, an FLRW spacetime has translational and rotational isometries. These take a particularly simple form if we decide to use the standard comoving coordinates: 
\begin{align}
\vec x & \to \vec x + \vec \epsilon \; , \qquad \epsilon = {\rm const} \\
\vec x & \to O \cdot \vec x \; ,  \qquad  O = {\rm const}, \quad O^T \cdot O = \mathbb{1} \; .
\end{align}
So, applying Coleman's argument, we can plug into the action the most general metric preserving the above symmetries: translational invariance in the form above means that nothing can depend on $\vec x$. Rotational invariance in the form above means that spatial indices can only come from $\delta_{ij}$ or $x^i$. So, the most general metric with these properties is
\begin{equation}
g_{00} = g_{00}(t) \equiv - N^2(t) \; , \qquad g_{0i} = 0 \; , \qquad g_{ij} = a^2(t) \delta_{ij} \; ,
\end{equation}
which is the standard minisuperspace ansatz. Coleman's argument guarantees that proceeding in this way is equivalent to looking for FLRW solutions at the level of the full Einstein equations.

The moral of this example is that, in order to apply Coleman's argument to a gauge theory, one should use an ansatz that is gauge-fixed enough so that all the symmetries one wants to preserve have a completely specified functional form, but not more. One should then use in the action the  most general ansatz compatible with such symmetries.

\section{Anisotropic cosmologies in minisuperspace}\label{minisuperspace}

Consider then the action for gravity and matter in  $d+1$ dimensions,
\begin{align} \label{EHA}
    S= \frac{1}{16 \pi G_N}\int d^{d+1} x \sqrt{-g } \ R \ + \ S_{\rm matter}\, .
\end{align}
We are interested in studying spatially flat, homogenous, but not necessarily isotropic solutions. ``Spatially flat, homogeneous" means that there are $d$ spacelike Killing vectors $\xi_{(1)}^\mu$, \dots, $\xi_{(d)}^\mu$ with the same algebra as $d$-dimensional Euclidean translations---that is, they commute.

Following the logic of the last section, we must fix the gauge enough so that the functional form of these Killing vectors is completely specified. Similarly to the FLRW case, one can show that one can always choose $x^1$, \dots, $x^d$ coordinates such that
\begin{equation}
\xi_{(i)} = \partial_i \; ,\qquad i = 1, \dots, d \; ,
\end{equation}
that is, such that $\xi_{(i)}^\mu$ generates rigid translations along $x^i$.

By requiring that our ansatz be invariant under these Killing vectors, we end up with a metric that does not depend on $\vec x = (x^1, \dots, x^d)$.  In ADM variables,
\begin{align} \label{adm}
  g_{\mu \nu} = \begin{pmatrix} -N^{2} + N^{k}N_{k} \ \ \ & N_{j}\\[3mm]
                            N_{i} & h_{i j}
                 \end{pmatrix} \, , \qquad
  g^{\mu \nu} = \begin{pmatrix} - \dfrac{1}{N^2} \ \ \ \ \ \ &\dfrac{N^j}{N^2} \\[4mm]
                \dfrac{N^i}{N^2} \ \ \ &  h^{i j}-\dfrac{N^i N^j }{N^2}
                \end{pmatrix} 
\end{align}
where $N$, $N^i$ and $h_{ij}$ are only functions of  time.  

Notice that under  a diffeomorphism of the form $x^i \rightarrow x^i + v^i(t)$, the shift transforms as 
$N^i \rightarrow N^i + \dot v^i\, $. So it is possible to get rid of $N^i(t)$ with a coordinate transformation. However, at face value this is incompatible with Coleman's argument: from the viewpoint of the variational problem, the only allowed truncations of configuration space are those that correspond to a symmetry requirement---that the ansatz preserve one or more of the symmetries of the action. Setting $N^i$ to zero is not a statement of symmetry, unless we were trying to impose isotropy as well, which we are not. We should thus keep $N^i$ around in the action, vary w.r.t~it to get the associated equation of motion (the momentum constraint), and only then set it to zero. However, as we show in Appendix~\ref{app_3}, for the matter systems that we will consider, the momentum constraint itself  guarantees that it is consistent to set $N^i$ to zero directly at the level of the action and not worry about the momentum constraint any further. We will then do so, and postpone justifying in detail this course of  action until the Appendix.

Notice that, after we set $N^i$ to zero by a change of coordinates, the new coordinates are still Killing coordinates, in the sense that the Killing vector fields are still of the form $\partial_i$. 
 The line element reduces to 
\begin{equation} \label{metricform}
ds^2 = - N^2(t) dt^2 + h_{ij}(t) dx^i d x^j\, .
\end{equation}
Contrary to the momentum constraint, the Hamiltonian constraint is nontrivial, and so
we have to keep the lapse $N$ as a degree of freedom in the action. However, we will systematically set $N=1$ in the equations of motion.

\subsection{Symmetries of the gravitational action}\label{symmetries}
For the ansatz \eqref{metricform}, the action becomes    
\begin{equation} \label{bianchiaction}
S= \frac{1}{16 \pi G_N}\int  \frac{dt }{4 N} \sqrt{h } \left(h^{il} h^{jm} - h^{ij} h^{lm}\right) \dot{h}_{ij} \dot{h}_{lm} + \ S_{\rm matter}\, .
\end{equation}
Notice that eq.~\eqref{metricform} does not correspond to a complete gauge-fixing: if we multiply $\vec x$ by a constant (in $\vec x$ and $t$) invertible matrix $L$, 
\begin{equation}
x^i \rightarrow \tilde x^i = L^i_{\ j} x^j \; ,
\end{equation}
the metric still takes the form  \eqref{metricform}, with the same $N(t)$, but with a new $h_{ij}(t)$:
\begin{equation} \label{transformations}
h_{ij} \rightarrow \tilde h_{ij} = (L^{-1})^k_{\ i} (L^{-1})^l_{\ j} \ h_{kl}\, .
\end{equation}
General (invertible) linear transformations in $d$ dimensions form the $d^2$-dimensional group $GL(d,\mathbb{R})$. In terms of its fundamental representation, it is useful to classify its generators $\ell_n$ in the following way:
\begin{enumerate}
\item {\em Dilations}: the associated generator is the identity matrix, 
\be
(\ell^{\, D}) {}^{ij} = \delta^{ij} \; ,
\ee
which rescales $h_{ij}$ by an overall constant. 

\item {\em Rotations}: the associated generators are antisymmetric matrices, for instance of the form 
\be \label{parametrize rotations}
(\ell^{\, R}_{ ab}) {}^{ij} =  \delta^i_a \delta^j_b -\delta^j_a \delta^i_b  \; ,
\ee
which generate rotations in the $(x^a,x^b)$ plane. We have $\frac{d(d-1)}{2}$ of them. The associated group, $SO(d)$, is the maximal subgroup of $SL(d, \mathbb{R})$.

\item {\em Diagonal shears}: the associated generators are diagonal, traceless matrices, for instance of the form
\be
\ell^{\rm \, diag} _{ a} = {\rm diag}(0 , \dots, 0, +1, -1, 0, \dots , 0 ) \; ,
\ee
where the $+1$ is at the $a$-th position along the diagonal, so that $\ell^{\rm \, diag} _{ a}$ generates shears along the $(x^a, x^{a+1})$ directions. We have $d-1$ of them.

\item {\em Off-diagonal shears}: the associated generators are fully off-diagonal, symmetric matrices, for instance
of the form
\be
(\ell^{\rm \, off} _{ ab}) {}^{ij} =  \delta^i_a \delta^j_b + \delta^j_a \delta^i_b \; , \qquad a \neq b\; ,
\ee
which generate shears along the $(x^a+x^b, x^a-x^b)$ directions.
We have $\frac{d(d-1)}{2}$ of them.
\end{enumerate}
Consider for now only the gravitational (i.e., Einstein-Hilbert) part of \eqref{bianchiaction}, and let's investigate how these generators act. Dilations rescale the spatial metric by a constant factor.  Despite their being a symmetry of the equations of motion, they are {\em not} a symmetry of the action \eqref{bianchiaction}, and so there is no conserved charge associated with them. The reason is that in going from~\eqref{EHA} to the reduced action~\eqref{bianchiaction}, we have integrated over the spatial volume. In other words, of the invariant combination $d^d x \sqrt{h}$, we are left only with the volume element $\sqrt{h}$.  As a result, the action is only invariant under special linear transformations $SL(d, \mathbb{R})$, that is, the subgroup of $GL(d,\mathbb{R})$ with unit determinant. 
All the other generators are symmetries of the Einstein-Hilbert part of the action \eqref{bianchiaction}, and so, neglecting matter for now, they correspond to conserved charges.

However, it turns out that for any given initial conditions, we can set to zero some of the charges by a suitable choice of coordinates, that is, by acting with the symmetry group itself. This is the  the same logic that allows one to reduce a point-particle mechanics problem with a central potential in three spatial dimensions to a two-dimensional one: since the angular momentum vector $\vec J$ is conserved, motion happens on a fixed plane; aligning the $x,y$ coordinates with such a plane corresponds to setting $J_x$ and $J_y$ to zero and keeping only a nonzero $J_z$.

For a generic symmetry group, there are arguments that indicate that one can always apply symmetry transformations so that the only charges that are turned on are those that make up the Cartan subalgebra, that is, the maximal abelian subgroup \cite{YY, BW}. We can confirm this general expectation in our specific case. We do so in Appendix \ref{charges}. For $SL(d, \mathbb{R})$, the Cartan subalgebra is spanned by the diagonal shears above. {\em If} the matter action is also invariant under $SL(d, \mathbb{R})$---a question that we will address soon---classical solutions will carry these $d-1$ conserved charges. If on the other hand the matter action is, for instance, only invariant under rotations, then the classical solutions will carry angular momentum charges, as many as the rank of $SO(d)$, which is the integer part of $d/2$. 

Below we will study in some detail the $d=2$ case, so let's see explicitly how things work out there.
Using the same numbering as in the general classification above, a convenient basis for the generators $GL(2,\mathbb{R})$ is 
\begin{equation} \label{generators}
\ell_1 = \begin{pmatrix} 1& 0\\
0 &1 \end{pmatrix}, \quad 
\ell_2 = \begin{pmatrix} 0& 1\\
-1 &0 \end{pmatrix}, \quad 
\ell_3 = \begin{pmatrix} 1& 0\\
0 &-1 \end{pmatrix}, \quad 
\ell_4 = \begin{pmatrix} 0& 1\\
1 &0 \end{pmatrix}\, , \qquad (d=2) \; .
\end{equation}

In agreement with our discussion above, upon exponentiation, $\ell_1$ generates a scale transformation, $\ell_2$ a rotation,
and $\ell_3$ and $\ell_4$ shears in the $(x^1,x^2)$ plane, respectively along the $x^1, x^2$ and $x^1 \pm x^2$ directions.  
It can also be useful to characterize the generators in terms of (real) irreducible representations of the $SO(2) \subset GL(2,\mathbb{R})$ generated by $\ell_2$: $\ell_1$ is pure trace (helicity zero), $\ell_2$ is antisymmetric (helicity zero), $\ell_3$ and $\ell_4$ are symmetric and traceless (helicity $\pm2$). In fact, under   a $45^\circ$ rotation, $\ell_3$ and $\ell_4$ transform into each other.

Consider now the most general time-dependent 2D spatial metric, $h_{ij}(t)$. It has three independent components, which we can parametrize conveniently as 
\begin{equation}\label{metric-rotation}
h_{ij} \ = A^2 \ \begin{pmatrix} \cos \frac{\theta}{2} & \ \ \sin \frac{\theta}{2} \\[2mm]
-\sin \frac{\theta}{2} & \ \ \cos \frac{\theta}{2} \end{pmatrix} \begin{pmatrix} e^{ \xi } & \ \ 0\\[2mm] 0& \ \ e^{- \xi} \end{pmatrix} \begin{pmatrix} \cos \frac{\theta}{2} & \ \ - \sin \frac{\theta}{2} \\[2mm]
\sin \frac{\theta}{2} & \ \ \cos \frac{\theta}{2} \end{pmatrix}\, ,
\end{equation}
where the common scale factor $A$, the shear $\xi$, and the rotation angle $\theta$ are all functions of  time. The factor of $2$ in the definition of $\theta$ is conventional, and is useful for what follows.
For the moment, suffice it to say that declaring  $\theta$ to be an angular variable of period $2\pi$ is consistent with the fact that a rotation of $\frac\theta2 = \pi$ is enough to map $h_{ij}$ into itself ($h_{ij}$ only has components of helicity zero and $\pm 2$).

The gravitational action in these new variables becomes
\begin{align} \label{action-theta}
       S_{\rm grav} &= \frac{1}{8\pi G_N}\int  \frac{dt }{N} \left( -\dot A ^2 + \frac14 A^2 \big(\dot \xi^2 +   \sinh^2 \! \xi \, \dot{\theta}^2 \big) \right) \; .
\end{align}
In App.~\ref{app-a} we give the explicit transformations of the fields, whose infinitesimal versions read,
\begin{align} 
\ell_2:& \qquad \quad \delta \theta = 1 , \qquad \qquad \qquad \quad \delta \xi = 0,\\
\ell_3:& \qquad \quad \delta \theta = - \sin \theta \coth  \xi , \qquad \delta \xi = \cos  \theta, \\
\ell_4:& \qquad \quad \delta \theta = + \cos  \theta \coth  \xi , \qquad \delta \xi = \sin  \theta\, ,
\end{align}
where we are suppressing the associated infinitesimal transformation parameters. 
One can check that the action~\eqref{action-theta} is invariant under these transformations. 
On the other hand, dilations simply rescale $A$ and, as we have already emphasized, they are {\em not} a symmetry of 
\eqref{action-theta}.

\subsection{Anisotropic cosmologies as trajectories in a meta-universe}
It is illuminating to think of our variables $A, \xi, \theta$ as coordinates $X^M$ in a $(2+1)$-dimensional meta-universe.
Then, the gravitational action \eqref{action-theta} is simply that of a free massless particle with trajectory $X^M(t)$ in so-called parametrized (or Polyakov) form, 
\be \label{free particle}
S_{\rm grav}  \propto \frac12 \int \frac{dt }{N} \, G_{MN} (X) \dot X^{M} \dot X^N \; ,
\ee
where $G_{MN}$ is the metric in this meta-universe:
\be
ds_{\rm meta}^2 = G_{MN}(X) dX^M dX^N = - d A^2 + \frac{A^2}{4} \big( d \xi^2 + \sinh^2 \! \xi \, d \theta^2 \big) \; .
\ee
Were it not for the relative factor of $4$, this would be the metric of flat space in Milne coordinates. However, with the relative factor of $4$, this is a more general (less symmetric) open FLRW cosmology. In fact, at fixed time-variable in the meta-universe, that is, at fixed $A$, this metric describes a two-dimensional hyperboloid, with $\xi$ and $\theta$ playing the roles of the standard radial and angular coordinates (this is another reason why the factor of 2 in \eqref{metric-rotation} is a convenient choice.) 
Related to this, notice that our symmetry group $SL(2,\mathbb{R}) \simeq SO(2,1)$ 
spanned by the three generators $l_2$, $l_3$, and $l_4$ is nothing but the isometry group of the hyperboloid, with $l_2$ being the rotation and $l_3$ and $l_4$ the generalized translations. 

Similar considerations apply in higher dimensions, albeit with less symmetric spatial sections in the meta-cosmology. To see this, let's start again from \eqref{bianchiaction} and let's decompose $h_{ij}$ into a common scale factor $A(t)$ and a unit-determinant matrix,
\be
h_{ij}(t) = A^2 (t) \, u_{ij}(t) \; , \qquad \det u =1 \; .
\ee
Taking into account the unit-determinant condition on $u$, the gravitational part of the action becomes
\be \label{Sgrav in d dim}
S_{\rm grav}= \frac{1}{16 \pi G}\int  \frac{dt }{N} \Big(-d(d-1) A^{d-2} \dot A^2 + A^d \, u^{il} u^{jm} \, \dot{u}_{ij} \dot{u}_{lm} \Big)\, .
\ee
Now, $u$ is a symmetric matrix with unit determinant, and so it has $\frac{d(d+1)}{2}-1$ independent components, exactly as many as the dimensions of the $SL(d, \mathbb{R})/SO(d)$ coset space. This is no accident: 
using the standard framework of nonlinear realizations \cite{CCWZ1,CCWZ2}, it is useful to think of the most general $u_{ij}(t)$ as being the result of applying a suitable $SL(d,\mathbb{R})$ transformation to the identity matrix. However, in the language of spontaneous symmetry breaking, the identity matrix breaks $SL(d,\mathbb{R})$ down to its $SO(d)$ subgroup, and so we can obtain a generic $u_{ij}(t)$ by acting on the identity with the {\em broken} generators only (type 3 and 4 in the classification above), for instance exponentiated as
\be
u_{ij}(t) = \big(e^{X^I(t) \, T_I} \big)^k {}_{i}  \big(e^{X^I(t) \, T_I} \big)^l {}_{j} \, \delta_{kl} = \big(e^{2 X^I(t) \, T_I} \big)_{ij} \; , 
\qquad I = 1, \dots,  \sfrac{d(d+1)}{2}-1 \; ,
\label{Goldstones}
\ee
where the $X^I(t)$ are a set of ``Goldstone fields",  and we used the fact that the broken $T_I$ generators in the fundamental representation correspond to symmetric matrices. The Goldstone fields can be thought of as coordinates on the coset $SL(d, \mathbb{R})/SO(d)$, and provide a parametrization of the most general positive definite, symmetric, unit-determinant $u_{ij}$. One can of course decide to put different coordinates on the coset space, and so from now on we can consider the $X^I$'s to be a generic set of coset coordinates. They are the higher dimensional analogs of our $\xi$ and $\theta$ variables above 
\footnote{In fact, our parametrization in \eqref{metric-rotation} is not the same as  \eqref{Goldstones} specialized to $d=2$, but is related to it by a nonlinear $(\xi, \theta) \leftrightarrow (X^1,X^2)$ change of variables.}.

Plugging the parametrization of $u_{ij}$ in terms of Goldstones into the gravitational action and defining the meta-cosmic time 
\be
X^0 \equiv A^{\frac{d}{2}} \; ,
\ee
we get
\be \label{S Goldstones}
S_{\rm grav}= \frac{1}{16 \pi G}\int  \frac{dt }{N} \Big(-4\sfrac{(d-1)}{d} (\dot X^0)^2 + \frac12 (X^0)^2 \, g_{IJ} (\vec X)\dot X^I \dot X^J \Big) \,
\ee
where $g_{IJ}$ is the metric of the $SL(d, \mathbb{R})/SO(d)$ coset manifold in the $X^I$ coordinate system, which is invariant under generic 
$SL(d, \mathbb{R})$ transformations
\footnote{The $ SL(d, \mathbb{R})$ isometries of the  coset manifold do not fix the overall normalization of $g_{IJ}$. The normalization chosen in \eqref{S Goldstones} matches what we had in $d=2$ if we declare  ``the" metric of the hyperbolic plane to be $d \xi^2 + \sinh^2 \xi \,  d \theta^2$.}.

Up to an overall constant, we thus have once again the action of a free massless particle, eq.~\eqref{free particle}, living in a meta-cosmology with metric 
\be
ds^2_{\rm meta} = -(dX^0)^2 + (X^0)^2 \sfrac{d}{8(d-1)} \, g_{IJ} (\vec X) dX^I dX^J \; .
\ee
The scale factor of this meta-cosmology is linear in the cosmic time $X^0$, and the spatial sections are invariant under 
$SL(d, \mathbb{R})$. However, since these sections are $(\sfrac{d(d+1)}{2}-1)$-dimensional, and $SL(d, \mathbb{R})$ only has $d^2-1$ generators,  for $d = 3$ and above we don't have enough isometries to make the spatial sections maximally symmetric. Their geometry, for generic $d$, is nicely reviewed in \cite{Richard}.

Adding matter to \eqref{Sgrav in d dim} will correspond to adding more fields and potentially breaking some of the symmetries that we have discussed. In the meta-universe picture, this will amount to adding dimensions to the meta-universe, and to breaking some of the isometries by adding potentials for some of our point-particle's coordinates---in which case our particle will not follow geodesics anymore. Let's look at a few examples.

\subsection{Time-dependent scalar matter}
The simplest possibility is to consider a time-dependent scalar field $\phi(t)$ as matter, which is consistent with the  isometry that we are trying to impose (spatial homogeneity). In fact, it is invariant under the full $SL(d,\mathbb{R})$ symmetry group we have been discussing at length. Considering for simplicity a free massless scalar in $d=2$, the action becomes
\begin{align} \label{scalar+grav}
S & = \int  \frac{dt }{N}  \bigg[\frac{1}{8\pi G} \left( -\dot A ^2 + \frac14 A^2 \big(\dot \xi^2 +   \sinh^2 \! \xi \, \dot{\theta}^2 \big) \right) + \frac12 A^2 \dot\phi^2 \bigg] \qquad \qquad
(d=2)  \nonumber \\
& = \frac{1}{8\pi G} \int  \frac{dt }{N}  \bigg[ -\dot A ^2 + \frac14 A^2 \big(\dot \xi^2 +   \sinh^2 \! \xi \, \dot{\theta}^2 + 2 \dot \phi_{\rm Pl}^2\big) \bigg] \; ,
\end{align}
where $\phi_{\rm Pl}$ is simply $\phi$ measured in Planck units.
As anticipated, we have effectively added one dimension to the meta-universe. Our fictitious particle still follows geodesics. Moreover, $SL(2, \mathbb{R})$ still corresponds to isometries of this enlarged meta-universe. In fact, we have a shift symmetry on $\phi_{\rm Pl}$ as well, and so we have one more isometry. As a consequence, any solution of the equations of motion will conserve all   $SL(2, \mathbb{R})$ charges as well as the shift-symmetry charge. 

Following the discussion of the preceding subsection, for any given initial conditions we can perform a symmetry transformation and set to zero all the charges that are not in the Cartan subalgebra. For $SL(2, \mathbb{R})$, this means that, up to coordinate transformation, only the charge associated with $\ell_3$ survives. Calling $Q$ the charge associated with $\phi_{\rm Pl}$-shifts and $Q_a$ the charges associated with the $\ell_a$'s of $SL(2, \mathbb{R})$, we have that, with suitable normalizations (and with $N=1$),
\begin{align}
&Q  = A^2 \dot \phi_{\rm Pl} \\ 
& Q_2  = A^2 \sinh^2 \! \xi \, \dot{\theta} \\
& Q_3  = A^2 \big(\dot \xi \cos \theta - \dot \theta \sin \theta \sinh \xi \cosh \xi \big)\\
& Q_4  = A^2 \big(\dot \xi \sin \theta + \dot \theta \cos \theta \sinh \xi \cosh \xi \big)
\end{align}
Setting $Q_2$ and $Q_4$ to zero while keeping $Q_3$ nonzero corresponds to setting 
\be
\theta = 0 \; , \qquad \dot \theta = 0 \; .
\ee
This can be thought of as an initial condition, but, by conservation of $Q_2$ and $Q_4$, it will be true at all times. Also, we emphasize again that {\em all} initial conditions can be put in this form by a suitable coordinate transformation.

We can thus just forget about $\theta$ altogether, and focus on the $A, \xi, \phi_{\rm Pl}$ degrees of freedom. Their dynamics are completely determined by conservation laws: the Hamiltonian constraint and the conservation of $Q$ and $Q_3$ read
\begin{align}
H^2(t) & =\frac14 \frac{Q_3^2 + 2 Q^2}{A^4(t)}  \; , \qquad H(t) \equiv \frac{\dot A}{A} \\
\dot \xi(t) & = \frac{Q_3}{A^2(t)} \\
\dot \phi_{\rm Pl}(t) & = \frac{Q}{A^2(t)} \; , 
\end{align}
with constant  $Q$ and $Q_3$.

And so, in particular, the contribution of anisotropy ($Q_3$) to the Friedmann equation scales like $A^{-4}$, and so does the energy density of  $\phi$. The anisotropy itself, measured as the difference in the Hubble rate along the two principal axes,
\be
\Delta H \equiv \frac{\frac{d}{dt}(A e^{\xi/2})}{ A e^{\xi/2}} - \frac{\frac{d}{dt}(A e^{-{\xi/2}})}{ A e^{-{\xi/2}}} = \dot \xi \; ,
\ee
scales as $A^{-2}$. 

We can generalize this analysis to $d > 2$ and to a scalar with a nonzero potential $V(\phi)$. We will have full $SL(d,\mathbb{R})$ symmetry, but no shift symmetry. By a judicious choice of coordinates, we can set to zero all the $SL(d,\mathbb{R})$ charges corresponding to rotations and to off-diagonal shears. We will thus be left with the $d-1$ charges associated with the diagonal shears. This means that, in this coordinates, the metric itself will be $A^2$ times a diagonal shear of the identity matrix, parametrized by $d-1$ ``anisotropy Goldstones" $\xi^a(t)$, for instance as
\be
h_{ij} = A^2 \, {\rm diag} \big( e^{\xi_1}, \, e^{\xi_2-\xi_1}, \dots , \, e^{\xi_{d-1}-\xi_{d-2}}, \, e^{-\xi_{d-1}} \big) \; .
\ee 
Since the generators in question commute, the coset submanifold spanned by these Goldstones is flat. So, for this reduced set of variables, up to factors the action must take the form (see eq.~\eqref{S Goldstones})
\be \label{dilationsandrest}
S \sim \int \frac{dt}{N} \big( -A^{d-2} \dot A^2  + A^d (\dot \xi^a \, {}^2 + \dot \phi^2_{\rm Pl}) - N^2 A^d \, V_{\rm Pl} (\phi_{\rm Pl})\big) \; ,
\ee
where $V_{\rm Pl}$ is the potential for $\phi$ in Planck units. The story then is very similar to the $d=2$ case, apart from the fact that now there is no shift-symmetry for $\phi$ and so both the dynamics of $\phi$ and those of the scale factor will be more complicated. Nonetheless, the dynamics of anisotropies is still completely determined by the conservation laws for the nonzero charges $Q_a$,
\be
\dot \xi^a (t) = \frac{Q_a}{A^d(t)} \; , \qquad Q_a = {\rm const} \; ,
\ee
and so is their contributions to the Friedmann equation, 
\be
H^2 \supset \frac1{A^{2d}(t)} \cdot \sum_a  Q^2_a  \; , \qquad Q_a = {\rm const} \; .
\ee

All this is in agreement with standard results, but here we derived it using only symmetries and conservation laws.

\subsection{Solid matter}\label{sec_solid}

We now consider what kind of matter can violate some of the $SL(d,\mathbb{R})$ symmetries of the gravitational action. We are particularly interested in the case in which the residual symmetries make up the $SO(d)$ rotation subgroup, so that solutions can be characterized in terms of conserved angular momentum charges---that is, they can be thought of as {\em rotating} solutions.

Ideally, we would like the matter action to be a function of $h_{ij}$ that is only invariant under $SO(d)$, for instance because it is a function of the trace of $h$, the trace of $h^2$, etc. However, precisely because such $SO(d)$ invariants are not invariant under the full $SL(d, \mathbb{R})$, they cannot arise directly as the mini-superspace limit of diff-invariant 
quantities---$SL(d,\mathbb{R})$ is what is left of diff-invariance in mini-superspace.

Following \cite{Dubovsky:2004sg}, the solution is clear: introduce scalar fields playing the role of St\"uckelberg fields, with  internal symmetries that have exactly the same algebra as the symmetries one wants to preserve (the $d$-dimensional Euclidean group, in our case), and then work in unitary gauge where these scalars are aligned with the coordinates. 
We now describe why, in our case, this is the same as considering a cosmological solid.

The low energy effective description of a solid or a fluid involves as many scalar fields $\phi^I$ as there are spatial dimensions ($d$), and a Lagrangian equipped with certain internal symmetries \cite{Soper:1976bb, DGNR}
\begin{align}
    \text{(internal translations = shifts)} \qquad \phi^I \longrightarrow & \ \ \phi + a^I, \qquad a^I = {\rm const} \label{phi shifts}\\
    \text{(internal rotations)} \qquad \phi^I \longrightarrow & \ \ O^I {}_J \phi^J, \qquad O^I {}_J \in SO(d) . \label{phi rot}
\end{align}
(We are for simplicity considering only solids that, at least at long distances, have isotropic ground states.)
On top of those, the fluid Lagrangian is also invariant under internal volume-preserving diffeomorphisms. A straightforward way to implement these symmetries to lowest order in derivatives is through certain invariants built out of
\be \label{B^IJ}
B^{IJ} \equiv g^{\mu\nu}\partial_{\mu} \phi^I  \partial_{\nu} \phi^J \; ,
\ee
which is a spacetime-scalar, shift-invariant symmetric matrix with internal indices.

Then, for a generic solid in $d$ spatial dimensions, the effective Lagrangian must be a function of the $d$ independent $SO(d)$ invariants that one can build out of $B$. For instance, following \cite{solidinflation}, one can take the trace of $B$ and normalized versions of the traces of $B^{n}$:
\be
X =  [B] \; , \qquad Y_2 = \frac{[B^2]}{[B]^2} \; \qquad \dots \qquad Y_d = \frac{[B^d]}{[B]^d} \; ,
\ee
where $[\,\cdots]$ denotes the trace of the matrix within.
To lowest order in derivatives, the action of a solid in $d+1$ dimensions then is
\be \label{solid action}
S_{\rm solid} = - \int dt \ h^{1/2} N F(X,Y_2, \dots, Y_d) \; ,
\ee
where $F$ is a generic function, related to the equation of state of the solid.

For our minisuperspace application, it is particularly convenient to work in so-called unitary gauge: recall that we are restricting to ansatze that are invariant under constant shifts  of the spatial coordinates $x^i$. If we consider the field configuration
\be \label{phi ground state}
\phi^I (\vec x, t) = x^I \; ,
\ee
we are formally breaking the spatial translations, but, given the shift symmetries \eqref{phi shifts}, we are in fact preserving a diagonal combination of internal shifts and spatial translations. This makes this $\phi^I$ configuration compatible with the mini-superspace framework.
Similarly, although spatial rotations are formally broken by the $\phi^I$ configuration above, thanks to the internal rotational symmetry \eqref{phi rot}, there is an unbroken diagonal combination of internal rotations and spatial ones. 
On the other hand, unless we demand that the solid action only be a function of $\det B$---in which case we would be describing a perfect {\em fluid} \cite{Soper:1976bb, DGNR}---the rest of $SL(d,\mathbb{R})$ is broken by \eqref{phi ground state}.
Notice that in unitary gauge, the matrix $B^{IJ}$ defined in \eqref{B^IJ} reduces simply to
\be
B^{IJ} = h^{IJ} \; .
\ee

The conclusion is that if we consider a generic solid in unitary gauge, we are effectively adding to the mini-superspace gravitational action a function of $h_{ij}(t)$ that is only invariant under the $SO(d)$ subgroup of $SL(d, \mathbb{R})$---precisely what we were looking for.

To make the discussion as concrete and as simple as possible, we now specialize to $d=2$. The $d=3$ case is analyzed in Appendix \ref{3D rotation}. Using the same parametrization of $h_{ij}$ as before, eq.~\eqref{metric-rotation}, we get that our invariants in unitary gauge read
\be \label{X and Y}
X = 2 A^{-2} \cosh \xi \;  , \qquad Y \equiv Y_2 = 1 - \frac{1}{2(\cosh \xi)^2} \; ,
\ee
which, consistently with the residual rotational invariance, depend on $A$ and $\xi$ but not on $\theta$. 
Since $F$ in \eqref{solid action} is a generic function of $X$ and $Y$, we can consider it directly as a generic function of $A$ and $\xi$, $F = F(A, \xi)$.
Notice that in the fluid case the matter action would only depend on  $\det B^{IJ} = A^{-4}$, \emph{i.e.}, $\partial_\xi F= 0$. In this limit the system would  still be invariant under the full $SL(2,\mathbb{R})$ group.

In summary, including the gravitational part \eqref{action-theta} as well, we consider the action 
\begin{align}
       S = \int dt \, A^2 \Big[ \frac1N \Big( - \frac{\dot A^2}{A^2} + \frac{\dot{\xi}^2}{4} + \frac{\dot{\theta}^2 \sinh^2{\xi}}{4} \Big) -  \ N  F(A,\xi) \Big] \, ,
\end{align}
where for simplicity we have chosen $8 \pi G = 1$ units.

Now the only symmetry of this action is $SO(2)$, acting as a constant shift of $\theta$. The associated charge is the angular momentum $J = \dot \theta \, A^2 \sinh^2 \! \xi$ (setting $N=1$),
and the dynamics of $\theta$ are completely determined by its conservation:
\be \label{eqq_theta}
\dot \theta = \frac{J}{A^2 \sinh^2 \! \xi}  \; ,\qquad J = {\rm const.}
\ee
Since we only have one symmetry generator---that associated with angular momentum---there is no way now to use the symmetries of the action to set the angular momentum to zero. This angular momentum is thus as physical as it can be, and a nonzero value for it can be taken to mean, by definition, that the universe is {\em rotating}.

Denoting from now on the derivatives of $F$ with respect to $A$ or $\xi$ by subscripts, the other independent equations are the Hamiltonian constraint and the $\xi$ equation of motion:
\begin{align}
\label{xi_eq_lnA}
    &4 H^2 = \dot{\xi}^2 + \frac{J^2}{A^4 \sinh^2 \! \xi} + 4 F  \\ 
    &\ddot{\xi} + 2H \dot{\xi} -  \frac{ J^2 \cosh \xi}{A^4 \sinh^3 \! \xi}   +  2 F_{\xi}  = 0
\end{align}
where $H \equiv \dot A/A$.

For both numerical and qualitative analyses, it proves useful to condense the above equations into a single second order one for $\xi ({\cal N})$, where ${\cal N} = \ln A$ is the number of $e$-folds and a prime denotes differentiation with respect to it:
\begin{multline}   \label{xinumerics}
\xi '' + 2 \xi '  - \frac{{\xi'}^3}{ 2 }- \frac{\xi'}{H^2} \left(\frac{J^2}{2 A^4 \sinh^2 \xi} - \frac{A F_A}{2}\right) 
- \frac{ J^2 \cosh \xi}{H^2 A^4 \sinh^3 \! \xi}  + \frac{2 F_{\xi}}{H^2} = 0\, .
\end{multline}
With this notation, $H^2$  stands for
\begin{equation} \label{hsq}
H^2 = \left(4 - {\xi'}^2 \right)^{-1} \left(\frac{J^2}{ A^4 \sinh^2 \! \xi}+4F\right)\, .
\end{equation}

\section{Understanding rotation}\label{understand}

Solutions with $J \neq 0$, or, equivalently, $\dot \theta \neq0$, are characterized by some ``rotation". But rotation of what with respect to what? 
We should note that the solid's volume elements, which evolve in time following trajectories with $\phi^I = {\rm const}$, in fact  follow geodesics. Indeed, $x^i = {\rm const}$ is always a geodesic for a metric of the form~\eqref{metricform}. Had we left the shifts $N_i$ (the $g_{0i}$ components of the metric) undetermined, the momentum constraint in the presence of a solid would enforce $N_i=0$ in unitary gauge (see Appendix \ref{app_3}). In other words, homogeneity of the cosmological solution seems to require that the solid's volume elements follow geodesics, as just stated. So there is no obvious sense in which the solid is rotating. One can also calculate the invariant vorticity associated with the solid's velocity field and easily check that it vanishes. 

By looking at~\eqref{metric-rotation}, it is rather clear that what is rotating here is the principal axes of expansion, with respect to the comoving frame $x^i = {\rm const}$. This is the case, at least, when $\dot \xi\neq 0$. However, our system can also support rotating solutions with $\dot \xi =0$. 

In more standard cases,  a constant $\xi$ solution simply corresponds to an FLRW universe ``in the wrong coordinates". That is, a constant $\xi$ can  be set to zero by a coordinate transformation, as a consequence of the $SL(2, {\mathbb R})$ symmetries. We can see this  explicitly in our formalism as well, if we specialize \eqref{xinumerics} to non-rotating ($J = 0$), constant $\xi$ ($\xi' = 0$) configurations.  We obtain the condition 
\begin{equation} \label{mycondition}
F_\xi (\xi, A)= 0 \, .
\end{equation}
Recall that now $A$ is effectively playing the role of time. So, the above condition must be obeyed at all $A$'s. {\em If} it is also obeyed for all values of $\xi$, then we are in the fluid limit, where the matter action only depends on the determinant of $B^{IJ}$, which in unitary gauge is $A^{-4}$.  
So, in this case we have an enhanced symmetry ($SL(2, {\mathbb R})$ vs.~$SO(2)$) that allows us to rescale $\xi$ to zero. 
This the standard case, that of a fluid-driven cosmology.\footnote{There is also the possibility that the condition \eqref{mycondition} is obeyed only at a specific $\xi = \xi_0$. In this case, the geometry is that of an FLRW cosmology, but the action for cosmological perturbations will likely be anisotropic.}

In the presence of rotation, $J \neq 0$, it is still possible to fine-tune the action so that there are solutions with constant $\xi$. This time, by looking at~\eqref{xinumerics}, we see that we need to impose the condition
\begin{equation}\label{justdroppedin} 
 \frac{J^2 \cosh \xi}{ A^4 \sinh^3 \! \xi} =  F_\xi \, .
\end{equation}
By choosing $F = A^{-4} f(\xi)$, the scale factor $A$ drops out of \eqref{justdroppedin} and the equation becomes a condition for $\xi$ only. 
The metric of the corresponding rotating solution is not that of an FLRW space, as one can check by explicitly calculating the spatial components of the Ricci tensor $R_{ij}$ and verifying that they are {\em not} proportional to the metric $h_{ij}$. We are in the presence of a genuinely rotating, anisotropic solution. The scale factor behaves like that of a kinetically dominated universe ($p/\rho =1$). We are away from the fluid limit and with just the  $SO(2)$  symmetry of the solid we cannot set the constant $\xi$ to zero.  Equivalently, the solid is in an anisotropic state $Y\neq1/2$. Loosely speaking, in this case the presence of angular momentum creates a centrifugal force that stabilizes the anisotropy at a certain constant value $\xi=\xi_0$.

\subsection{A simple model for a solid}
As an explicit example to use for a more in-depth study, we now specialize to the following solid Lagrangian, 
\begin{equation} \label{solid_model}
    F(X, Y) = X^{\epsilon}\, .
\end{equation}
This simple model generates, in the limit of zero shear and rotation ($\xi=\theta=0$), inflation with constant slow-roll parameter $\epsilon \equiv -\dot H/H^2$  \cite{solidinflation}. 
Equivalently, on an FLRW background, \eqref{solid_model} enforces the constant equation of state $p/\rho = \epsilon -1$. 
Using \eqref{X and Y}, eq.~\eqref{xinumerics} reduces to
\begin{align}  \label{numerics-2}
\xi '' - \xi ' (\epsilon-2)  + \frac{{\xi'}^3 (\epsilon -2)}{ 4 } & + \frac{\epsilon(4 - {\xi'}^2) \tanh\xi}{2}  \\
& + \frac{ J^2 \ e^{- 4 {\cal N}}}{4 H^2  \sinh^2 \xi}\left[(\epsilon - 2) \xi' - \frac{4}{\tanh \xi}\right] = 0 \, . \nonumber
\end{align}

This equation can easily be solved numerically by substituting~\eqref{hsq} in the second line. The qualitative behavior of solutions can also be captured in different limits, as we now show. 

First, one should notice that the last term proportional to $J^2$ becomes very rapidly subdominant with respect to the other terms. So, it can be ignored  to a very good approximation, and one can concentrate on just the first line of~\eqref{numerics-2}. We will do so from now on.

Second, after trial and error one discovers that the equation is more easily studied in the variable
\be
y({\cal N}) \equiv \log (\sinh \xi({\cal N})) \; ,
\ee
in terms of which the $\xi$ eom reduces to
\be \label{unified}
e^{2 y}\Big[ y'' -\sfrac{(2- \epsilon )}{4} y' \, ^3  -  \sfrac{\epsilon}{2} y' \,^2  + (2- \epsilon) y' + 2 \epsilon\Big]
+ y'' + y' \, ^2 + (2-\epsilon) y' + 2 \epsilon = 0 \; . 
\ee
Despite this  equation's looking complicated, it is quite easy to solve when $y $ is large in absolute value, thanks to the exponential factor in front of the bracket: for large positive $y$, the exponential is very large and the equation essentially reduces to the vanishing of the combination inside the brackets. For large negative $y$, the exponential is very small and the combination inside the bracket can be ignored. In both cases, the equation reduces to a {\em first}-order ODE for $y'$---$y$ itself disappears from the equation. 

\begin{figure}[H]
    \centering
 \includegraphics[]{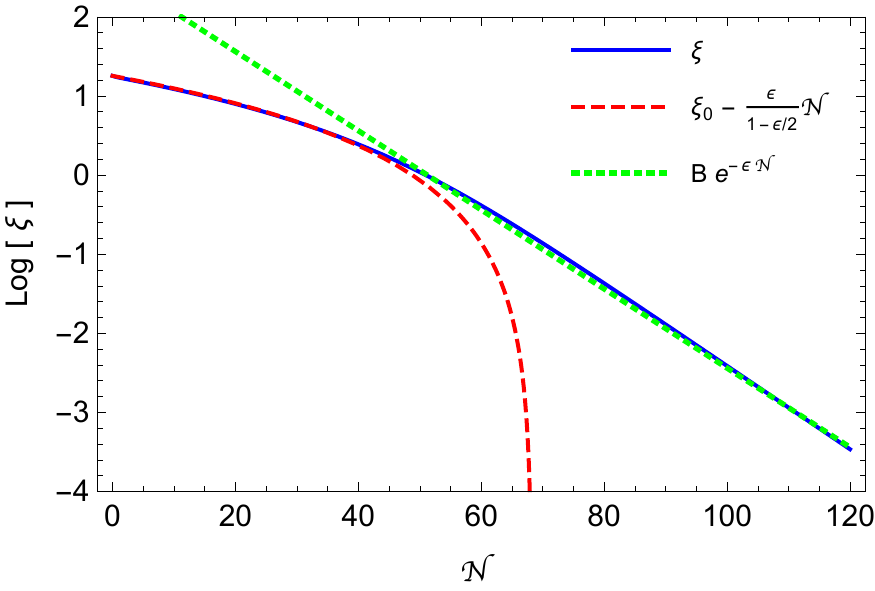}
    \caption{We numerically integrate the system for a sufficient number of $e$-folds ${\cal N}$ so that $\xi$ evolves from $\xi \gg 1$ to $\xi \ll 1$. As expected, the large-anisotropy regime is well approximated by \eqref{3.10}, while the low-anisotropy regime by the dominant mode in \eqref{xi_sol_small}, which we have approximated at small $\epsilon$ with $\sim e^{-\epsilon \cal{N}}$. Here $\epsilon = 0.05$. 
    }
    \label{xi_Sinhxi}
\end{figure}

It so happens that  in both limits the least decaying attractor solutions as ${\cal N}$ increases evolve  {\em linearly} in ${\cal N}$, with only slightly different rates,\footnote{At $\xi \gg1$ we find also the two behaviors $y'({\cal N}) = \pm 2$. The growing mode corresponds to the vanishing of one of the two principal expansion rates. This is a  repeller from the point of view of dynamical analysis. All configurations with $y'({\cal N}) > 2$ are unstable and have expansion rates of opposite signs.}
\begin{align}\label{3.10}
y'(\cal N)  & \simeq - \frac{\epsilon}{1- \epsilon/2}  = - \epsilon + {\cal O}(\epsilon^2) \; ,  & y \to + \infty \quad (\xi \gg 1) \; \; \\
y'(\cal N)  &  \simeq \epsilon/2 - \big( 1- \sqrt{1-3 \epsilon + \epsilon^2/4} \big) = - \epsilon + {\cal O}(\epsilon^2)  \; , & y \to -\infty \quad (\xi \ll 1) \; . \label{xi_sol_small}
\end{align}
So, keeping only the leading order in $\epsilon$, the least decaying solution for $\xi({\cal N})$ has
\be
\sinh \xi \sim e^{-\epsilon {\cal N}}   \qquad \mbox{for } \xi \gg 1  \mbox{ or } \xi \ll 1 \; ,
\ee
with a transient behavior around $\xi \sim 1$ ($\, |y| \lesssim 1$), which our method is not able to resolve. 
Fig.~\ref{xi_Sinhxi} shows the validity of our approximation for different values of $\epsilon$.

Once the approximated solutions for $\xi$ are found, we can deduce $\theta$ by integrating equation \eqref{eqq_theta}. By inspection, when the anisotropy has become small, $\xi \ll 1$, we obtain the asymptotic behavior  
\begin{equation} 
\theta ' \sim e^{-2 (1+\epsilon){\cal N}},
\end{equation}
with $\theta$ approaching its asymptotic value exponentially fast.

\subsection{Energy densities}
It is useful to interpret the terms in the Friedmann equation~\eqref{xi_eq_lnA} as the energy densities associated with 
an isotropic background ($\xi = \dot \theta = 0$), with anisotropy, and with rotation, 
\begin{align}
\rho_{\rm bkg} & = F|_{\xi = 0} \\
\rho_{\rm aniso} & = \ \rho_{\xi, {\rm kin}} + \rho_{\xi, {\rm pot}} \equiv \ \frac{\dot \xi^2}{4} + \big (F-F|_{\xi = 0} \big) \\  
\rho_{\rm rot} & =  \frac{J^2}{4 A^4 \sinh^2 \! \xi}\; ,
\end{align}
hopefully with obvious notation.
If we consider again our simple model \eqref{solid_model} and work for simplicity to lowest-order in $\epsilon$
we can  deduce the behaviors of these energy densities by looking at our explicit solutions.

Let's start with the isotropic background. It has
\be \label{rho bkg}
\rho_{\rm bkg} \sim e^{-2 \epsilon {\cal N}} \; .
\ee
As a consistency check, notice that if this were the only contribution to the Friedmann equation, we would have $H \sim e^{- \epsilon {\cal N}}$, which corresponds to exactly constant $-\dot H/H^2 = \epsilon$, as advertised above for the model \eqref{solid_model}.

\begin{figure}[H]
    \centering
    \includegraphics[width=7.5cm]{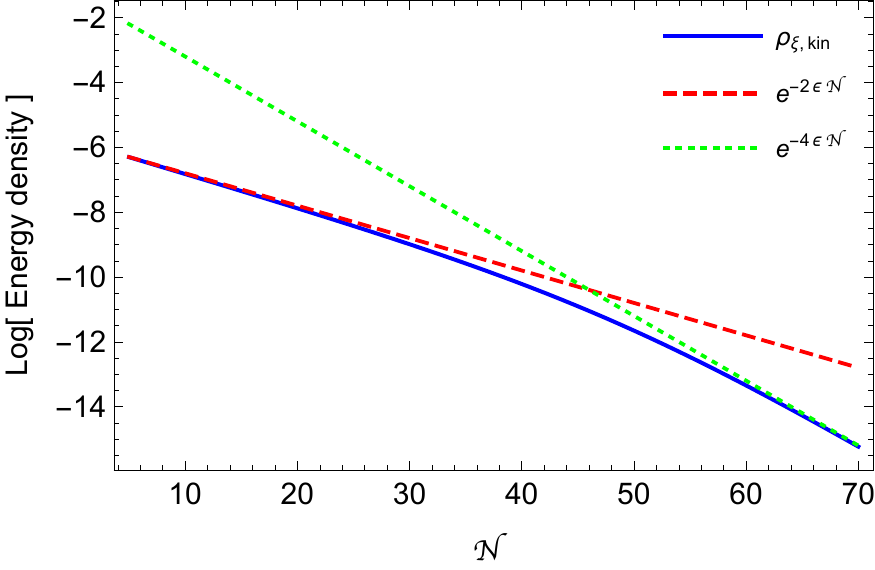}
    \includegraphics[width=7.5cm]{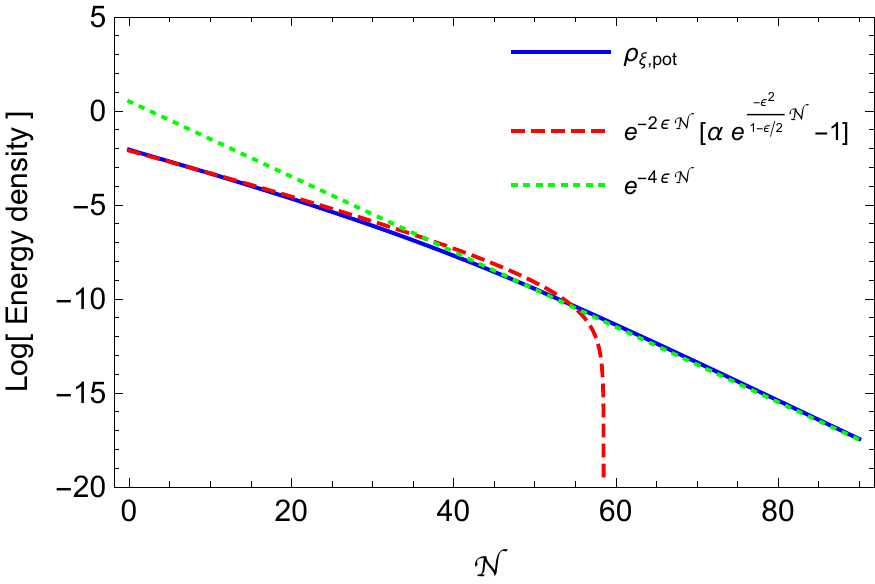}
    \caption{On the left, the kinetic energy density $\rho_{\xi, {\rm kin}}$ as a function of the number of e-folds ${\cal N} = \ln A$, and on the right, the potential energy density $\rho_{\xi, {\rm pot}}$, for $\epsilon = 0.05$. As expected from the approximate solutions \eqref{3.10} and \eqref{xi_sol_small}, there are two distinct modes corresponding  to high and small values  of $\xi$.
    } 
    \label{rho_xi_kin_tot}
\end{figure}

We can now assess the importance of anisotropy and rotation by comparing the behaviors of their energy densities to that of the background. 
For anisotropy, notice that $\xi' {}^2$ is well approximated by a constant at $\xi \gg 1$ and goes as $e^{- 2 \epsilon {\cal N}}$ when $\xi \ll 1$ (eqs.~\ref{3.10} and~\ref{xi_sol_small}). As a consequence, the kinetic part of $\rho_{\rm aniso}$ goes through two different regimes (Fig.~\ref{rho_xi_kin_tot}, left panel), 
\begin{equation}
\rho_{\xi, {\rm kin}} \sim e^{- 2 \epsilon {\cal N}} \quad (\xi \gg 1) \; , \qquad \qquad \rho_{\xi, {\rm kin}} \sim e^{- 4 \epsilon {\cal N}} \quad (\xi \ll 1) \, .
\end{equation}
The potential part of $\rho_{\rm aniso}$ also has two distinct modes depending on the anisotropy regime. For small $\epsilon \ll 1$,

\begin{equation}
\rho_{\xi, {\rm pot}} \sim e^{- 2 \epsilon {\cal N}}\left(\alpha e^{- \frac{\epsilon^2}{1-\epsilon/2} {\cal N}} -1 \right) \quad (\xi \gg 1) \; , \qquad  \qquad  \rho_{\xi, {\rm pot}} \sim e^{- 4 \epsilon {\cal N}} \quad (\xi \ll 1) \, ,
\end{equation}
with $\alpha$ some constant (see Fig.~\ref{rho_xi_kin_tot}, right panel).
So, as far anisotropies go, overall we have

\begin{equation}
\rho_{\xi, {\rm aniso}} \sim e^{- 2 \epsilon {\cal N}} \quad (\xi \gg 1) \; , \qquad \qquad \rho_{\xi, {\rm aniso}} \sim e^{- 4 \epsilon {\cal N}} \quad (\xi \ll 1) \, .
\end{equation}

\begin{figure}[H]
    \centering
    \includegraphics[width=8cm]{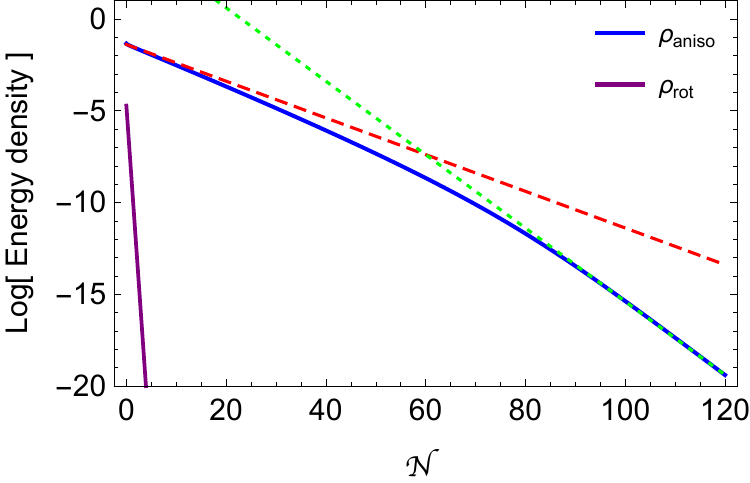}
    \caption{The energy densities of the anisotropy $\rho_{{\rm aniso}}$ and the rotation $\rho_{{\rm rot}}$, for $\epsilon =0.05 $ . The red and green dashed lines correspond resp. to the $e^{- 2 \epsilon \cal{N}} $ and $ e^{- 4 \epsilon \cal{N}}$ modes. }
    \label{rho_theta_vs_xi}
\end{figure}

Then, the energy density associated with rotations behaves as
\begin{equation}\label{smallepsilon}
\rho_{\rm rot}   \sim e^{-(4 - 2 \epsilon){\cal N}}\, .
\end{equation}
We see that, in terms of energy densities, the  contribution coming from  rotation decays much faster than that coming from  anisotropy (Fig.~\ref{rho_theta_vs_xi}); 
however, the latter scales precisely as the isotropic background's energy density in the high-anisotropy regime, eq.~\eqref{rho bkg}, and decreases slightly faster in the low-anisotropy regime. 
 In this simple model, and in stark contrast with more standard cosmological models with vanishing anisotropic stresses, the importance of anisotropies gets diluted very slowly by the Hubble expansion. This is in qualitative agreement with the conclusion of \cite{Bartolo:2013msa, Bartolo:2014xfa}, with a small quantitative discrepancy likely due to our working in $2+1$ dimensions.

\section{Attempts at a quantum theory}\label{quantum2}

Among the original motivations of this work was a better understanding of the two main pillars of cosmology---homogeneity and isotropy---in a fully quantum mechanical sense. Although somewhat detached from the main line of the paper, we report some of our (persisting) confusion on the subject in the following subsection. 
Then we concentrate on the much more modest task of quantizing the mini superspace model that we are dealing with classically. This is a relatively simple quantum system, whose ``diagonal version" (i.e. without rotation) has been throughly studied in the literature  (see e.g.~\cite{Christodoulakis:2018swq} and references therein). It seems to us that, even at this mini-superspace level, the rules of the game are not completely clear. The domain of validity of this approximation is not obvious, and neither is whether the corresponding truncation of degrees of freedom makes sense  quantum mechanically.

\subsection{Spacetime symmetries in quantum gravity}

It is hard to overestimate the role of spacetime symmetries. In particle physics Poincar\'e invariance is nothing less than foundational. In cosmology an equally important role is played by homogeneity and isotropy. However, when dynamical gravity if fully taken into account, we do not seem to have a grasp of what spacetime symmetries even mean: 
a given classical metric can have a number of isometries, but how do we characterize the spacetime symmetries of a wavefunctional of metrics? One could argue that when (quantum) gravity is dynamical, spacetime symmetries are simply lost. But this conclusion is probably too rash. Presumably, there \emph{is} a state of quantum gravity that corresponds to empty Minkowski spacetime and enjoys the symmetries of the Poincar\'e group. Also,  perturbative calculations indicate that the quantum state of our universe exiting primordial inflation is homogeneous and isotropic to an extremely high degree. Is there any non-perturbative (\emph{i.e.}, beyond cosmological perturbation theory) characterization of this?

Part of the confusion stems from the fact  that in quantum mechanics symmetry is a property of the wavefunction. 
This represents an interesting twist on classical physics. Let us consider the simple example of a scalar field $\phi$ in  
Minkowski space.  The state of such a field is \emph{homogeneous} if the wavefunction  $\Psi$ satisfies\footnote{Equivalently, and perhaps more famously, any $n-$point correlator of the field $\langle \phi(\vec x_1)  \phi(\vec x_2) \dots  \phi(\vec x_n)\rangle$
should only depend on the mutual distances among the points $\vec x_1, \dots , \vec x_n$, \emph{i.e.} on their intrinsic geometry, and not on their overall position.}
\begin{equation} \label{def1}
\Psi \big [\phi(\vec x \, ) \big]\ = \ \Psi \big[  \phi(\vec x - \vec a \, )  \big]\, ,
\end{equation}
for every vector $\vec a$. Notice that classical homogeneity (i.e., simply,  $\phi(\vec x\, ) = {\rm const.}$) has little to do with the above statement. In classical field theory restricting to configurations with certain spacetime symmetries implies a massive truncation of the phase space. No such a truncation is implied by~\eqref{def1}. 
 As far as we know, classically homogeneous configurations of the type $\phi(\vec x) = {\rm const.}$ could even be absent from the quantum ensemble~\eqref{def1}. In other words, $\Psi$ could vanish on each and every homogeneous configuration, and still be homogeneous! 

This is all well understood in the case of quantum fields on a given classical spacetime.
But what is the analogue of~\eqref{def1} for gravity?  We should be able to pose this question---albeit not necessarily to answer it---directly at the level of the low-energy effective theory. In canonical gravity a state is a functional of the three-dimensional metric $h_{ij}(\vec x \, )$ and of the matter fields. The property displayed in eq.~\eqref{def1} is clearly meaningless when applied to the metric field itself, because there is no such notion of ``translated metric $h_{ij} (\vec x - \vec a)$" in the absence of a background classical spacetime: the $\vec x$ coordinates are arbitrary, and the momentum constraint ensures that the wavefunction only depends on invariant quantities. So, how do we characterize translations for a generic metric (with no isometries) in a coordinate independent fashion?

While for translations we have nothing to say at the moment, for rotations we can use the mini-superspace approach that we have been using so far in the paper, and try to make sense of it quantum-mechanically. 
As we have seen, mini-superspace corresponds to a truncation of degrees of freedom already in the classical configuration space.  As we emphasized in Sec.~\ref{rules}, at the level of the classical variational problem this is a consistent truncation. At the quantum level, however, the question is more subtle: the degrees of freedom that we don't keep still have their quantum ``life", which manifests itself in nontrivial correlation functions/variances, even if we assume that those degrees of freedom are in their vacuum. From the modern viewpoint, the sensible question seems to be whether the minisuperspace action (perhaps with renormalized couplings) corresponds to a consistent low-energy theory where all non-homogeneous degrees of freedom have been integrated out. 

The answer appears to be `no'. For a spatially infinite universe, the modes that are not constant in $\vec x$ form a gapless continuum, and there is no hope of ending up with a local derivative expansion upon integrating them out. For a spatially finite universe, say compactified on a torus, the Kaluza-Klein (KK) modes are gapped, and in principle one can integrate them out. However, one would like the physical size of the universe to be larger than the Hubble radius, in which case the KK are lighter than the Hubble scale, and cannot be ignored during the evolution of the universe---in particular, one can have cosmological KK particle production.

In conclusion, as far as quantum cosmology goes, the minisuperspace approach should not be thought of as a consistent truncation of the full theory in the same sense as low-energy effective field theory is. Rather, it should be thought of as a toy model that drastically simplifies the theory while retaining some of the puzzling difficulties associated with gauge invariance.

With these qualifications in mind, we now proceed to consider the quantum mechanics of our models (and more general ones) in mini-superspace.

\subsection{Mini-superspace approach}

The gravitational part of the Hamiltonian in $d+1$ dimensions reads 
\begin{equation}    \label{hamiltonian}
H_{\rm grav} =  N\ h^{-1/2}  \left( \pi^{ij} \pi_{ij} -\frac{\pi ^2}{d-1}  \right)  \, ,
\end{equation}
where $\pi^{ij}$ is the momentum conjugate to $h_{ij}$, and indices are lowered by using $h_{ij}$ itself. To the Hamiltonian above, we should add the matter part, which we will discuss below.

As is well known, gravity is a constrained system where the only equations governing the wave function are constraints. Variation with respect to $N$ gives the Hamiltonian constraint, or Wheeler DeWitt (WdW) equation. When applied to the wavefuntion,
\begin{equation}
H \Psi = 0\, ,
\end{equation}
it enforces invariance under time reparameterizations. 

One may wonder what happened to the momentum constraints, which are normally obtained by varying with respect to the shifts $N_i$. In the classical theory, when one loses some constraints because of a gauge choice implemented directly in the action, one should make sure to impose the corresponding equations independently. 
Momentum constraints represent the invariance of the wave function under spatial diffeomorphisms. As previously mentioned, the remnant of this GR symmetry here is the (finite dimensional) special linear group. In discussing classical solutions for cosmologies driven by a time-dependent scalar field, so that there is no anisotropic stress, we have used this group to set the angular momentum to zero. Can we thus interpret $SL(d, \mathbb{R})$ as a gauge symmetry? 
Should we impose ``by hand" that $\Psi(h_{ij})$ be invariant under $SL(d, \mathbb{R})$? This sounds like the correct mini-superspace analog of the momentum constraint.

However, it is not difficult to see that this would kill the theory altogether. For example, in $2+1$ dimensions, the configuration space is the two dimensional hyperboloid and $SL(2, \mathbb{R}) \simeq SO(2,1)$ can take a point on the configuration space into any other point. Imposing the invariance of $\Psi$ under $SL(2, \mathbb{R})$ would mean to set it to a constant on the entire hyperboloid. Apart from obvious normalization problems, this would mean dealing with a Hilbert space of just one state. One might be temped to impose that $\Psi$
be invariant {only} under a compact subgroup of $SL(2, \mathbb{R})$---the rotations on the hyperboloid around the origin. This would solve the normalization problem and would leave some non-trivial dynamics to the system. But why should we restrict to this subgroup? 
We conclude that, although $SL(d, \mathbb{R})$ is morally the remnant of the full spatial gauge invariance of GR, we cannot interpret it as a gauge symmetry at the mini-superspace level.

 In the case of a solid-driven cosmology, things appear to be better defined. working in unitary gauge we  only have a residual $SO(d)$ invariance.  Moreover, as far as spatial diffeomorphisms go, unitary gauge is a complete gauge-fixing. So, unless we decide to interpret the internal rotational symmetry \eqref{phi rot} also as a gauge symmetry, we can now use the residual $SO(d)$ as a physical symmetry, and classify the physical states in terms of the associated angular momentum, without restricting only to states that are annihilated by it. 
It is then tempting to declare that the states that are {\em isotropic} in the quantum mechanical sense are those with zero angular momentum, since they are invariant under $SO(d)$. However, among these, there are those whose wave-function is peaked around isotropic cosmologies ($\xi = 0$, in the $d=2$ case), and those whose wavefunction is perhaps peaked around some highly anisotropic cosmology, but for which the anisotropy direction has been averaged over. These are two very different cases, and only the former  seems to be the quantum analog of an isotropic cosmology.
Clearly, at the moment we only have a  partial understanding of the quantum theory.

\subsection{Ordering etc.}

The WdW operator is the quantum version of the gravitational Hamiltonian \eqref{hamiltonian}, plus the matter part. 
Since $h_{ij}$ and $\pi^{ij}$, as quantum operators, don't commute, the expression \eqref{hamiltonian} is ambiguous at the quantum level. 

Using the wave-function representation of the quantum state, $\Psi = \Psi(h_{ij})$, the momentum conjugate
to $h_{ij}$ acts as a derivative:
\be
\pi^{ij} \ \longrightarrow \ -i\frac{ \partial}{\partial h_{ij}}\, .
\ee
So the question is: how are we to order the $h$ fields and the derivatives with respect to them when \eqref{hamiltonian} acts on the wave function?

The most reasonable option seems to be that of building the \emph{Laplace-Beltrami} operator of the corresponding superspace metric~\cite{DeWitt:1967yk,Hawking:1985bk}. In other words, whenever we are in the presence of a system (e.g. gravity + matter) whose Lagrangian/Hamiltonian can be cast in the form (see eq.~\eqref{free particle})
\be 
L =  \frac{1 }{2 N} \, G_{MN} (X) \dot X^{M} \dot X^N - N V(X) \ \ \Longrightarrow \ \ H = N\left(\frac{1}{2} \ G^{MN}\pi_M \pi_N + V \right)\, ,
\ee
it is natural to identify the kinetic part of the Hamiltonian with the Laplacian associated with the metric $G_{MN}$, 
\begin{equation}
 \label{laplacian}
    \nabla^2 = \frac{1}{\sqrt{-G}} \ \frac{\partial}{\partial X_M} \ \sqrt{-G} \ G^{MN} \frac{\partial}{\partial X_M} \, .
\end{equation}
In the above expression the determinant of $G_{MN}$ has been assumed to be negative because  the conformal mode of the metric is a ``time-like" direction in superspace (see e.g. eq.~\eqref{dilationsandrest}).

This choice of ordering has important advantages. It gives an operator that is Hermitian with respect to the ``standard" scalar product 
\begin{equation} \label{probability}
(\psi_1,\psi_2) = \int d^n X \sqrt{-G}\  \psi_1^*(X) \psi_2(X)\, ,
\end{equation}
once appropriate boundary conditions are chosen.\footnote{In order to interpret the WdW equation as a ``time-evolution",  one should use one of the fields, say, $X^0$, as a time variable and think of $\psi(X^0, X^1,\dots)$ as a probability amplitude for the remaining variables, ``evolving" in the time $X^0$. This approach would suggest a different scalar product than~\eqref{probability}, the so-called ``Klein-Gordon" scalar product~\cite{DeWitt:1967yk,Vilenkin:1986cy}. In this case the probability for the remaining fields $X^1, X^2, \dots$  is guaranteed to be conserved in ``time" for all solutions of the  WdW equation.  }
At the same time, and more importantly, the field-space Laplacian~\eqref{laplacian} is a well-defined prescription, invariant under field redefinitions. 
Other options (e.g. ``push all derivatives to the right" $\sim G^{MN} \partial_M \partial_N$) are clearly dependent on a specific choice of coordinates in field space and therefore cannot be considered as valid alternatives.

\subsection{(In-)consistent truncations}

One puzzling, potentially interesting, property of the choice~\eqref{laplacian} is that it depends on the dimensions of the meta-universe, in the same way as the radial part of the standard Laplacian is different if we work, say, in two rather than three dimensions. 
Take, for example, the Hamiltonian~\eqref{hamiltonian} in 2+1 spacetime dimensions.  The corresponding Laplacian operator reads (see App.~\ref{quantum} for details) 
\begin{equation} \nabla^2 = \frac{1}{\sqrt{h}} \left[ \left( h_{ik}h_{jl} -h_{ij}h_{kl} \right)\frac{\partial}{\partial h_{ij}}  \frac{\partial}{\partial h_{kl}} + \frac{1}{4} h_{ij} \frac{\partial}{\partial h_{ij}}\right]\qquad \rm (pure\ gravity) \, .\label{PG}
\end{equation}
On the other hand, if we add a standard scalar field as in~\eqref{scalar+grav} we get  
\begin{align} \label{lap-scalar} 
  \ \   \nabla^2_s =  \frac{1}{\sqrt{h}} \left[ \left( h_{ik}h_{jl} -h_{ij}h_{kl} \right)\frac{\partial}{\partial h_{ij}}  \frac{\partial}{\partial h_{kl}} \right] +   \frac{2}{\sqrt{h}}  \frac{\partial^2}{\partial {\phi}^2} \qquad \rm (gravity \, + \, scalar) .
\end{align}
The point here is that adding a scalar field also modifies the piece of the Laplacian containing derivatives with respect to the metric. 

Something similar happens if we change the number of degrees of freedom that we decide to keep in the metric itself. For instance, we can write the  operator~\eqref{lap-scalar} more explicitly in the $A$, $\xi$ and $\theta$ variables introduced in our general parametrization of a $d =2$ spatial metric, eq.~\eqref{metric-rotation}:  
\begin{align} \label{not-so-mini}
 \nabla^2_s = \frac{1}{2 A^2 } \left[-\frac{A^2}{4} \partial_A^2  -\frac{3}{4}A \ \partial_A   + \partial_{\xi}^2   +\frac{1}{ \tanh \xi} \partial_{\xi}     +\frac{1}{\sinh^2 \! \xi}  \partial_{\theta}^2  +  \partial_{\phi}^2 \right] \, .
\end{align}
However, if we consider an isotropic FLRW space instead of our Bianchi-type model, then from~\eqref{lap-scalar}, we get
\begin{align} \label{mini-super}
  \nabla^2_{s, {\rm FLRW}}= \frac{1}{2 A^2}\left[ -\frac{A^2}{4} \partial_A^2 -\frac14 A \ \partial_A  +   \partial_{\phi}^2 \right] \qquad \rm (gravity + scalar, FLRW)\, .
\end{align}

Because classical FLRW solutions are special cases of Bianchi models,
one might have hoped to recover~\eqref{mini-super} in some isotropic limit of~\eqref{not-so-mini}---for instance when the state $\Psi$ does not depend on $\xi$ or $\theta$. But this is clearly not the case, because~\eqref{mini-super} and~\eqref{not-so-mini} imply a different dependence of the wavefunction on $A$.  In the structure of the Laplacian, first derivatives with respect to $A$ are sensitive to the total number of fields involved, or even just on the number of symmetries that one wants to impose on the configuration space before quantization. In the beginning of this section we discussed how the  mini-superspace approach does not seem  a consistent truncation of the full theory. As an extreme example of this, we have just shown that the FLRW mini-superspace is not even a consistent truncation of a Bianchi one!  

\section{Conclusions}

In this paper we have highlighted an aspect of homogeneous cosmology, \emph{rotation}, which is associated with anisotropy and with the way its main axes evolve in time. The effect is physical only in the presence of  anisotropic stress in the energy momentum tensor. We have studied in some detail the case of solid matter, but it would be interesting to extend the analysis to other media. In particular, we have considered homogeneous and isotropic solids, characterized by a group of symmetry made up of internal translations and rotations [eqs.~\eqref{phi shifts} and~\eqref{phi rot}]. As discussed in Sec.~\ref{sec_solid}, this corresponds to breaking the  $SL(d,\mathbb{R})$  symmetry of the gravitational action down to $SO(d)$. However, it would be interesting to consider also less symmetric solids that are not fully invariant under internal rotations (e.g.~some ``crystal" or ``quasi-crystal"), for instance along the lines of \cite{Kang:2015uha}. In 2+1 dimensions, this would correspond to a matter Lagrangian that, in unitary gauge, explicitly depends on $\theta$ and not only on $A$ and $\xi$ (these parameters are defined in Sec.~\ref{symmetries}).  

Rotation could be a feature of some relevance in other inflationary models beside solid inflation, such as chromo-natural~\cite{Adshead:2012kp} and gaugid~\cite{Piazza:2017bsd} inflation. More generally, there could be traces of rotation even in more recent cosmic epochs, whenever the matter content of the universe features anisotropic stress. Although rotation is generally expected to rapidly dilute away like in the solid model considered in this paper, this does not need to be always case. 

Finally, we have pointed to some questions/puzzles of the quantum mechanical treatment. Among them, the structure of the Laplacian in field space  depends on the total number of fields involved. This directly affects the dependence of the wave function on the scale factor. One might object that we are not forced to use the Laplace-Beltrami operator as an ordering prescription for the Laplacian.  However, it is hard to conceive other equally sensible choices, invariant under field redefinitions. 

It is tempting to wonder whether, in a more realistic scenario, also long-wavelength, super-Hubble modes should be included in the degrees of freedom that affect the behavior of the scale factor. More generally, we would like to better understand how to operate a consistent truncation of degrees of freedom in quantum cosmology, in the same way we do in standard effective field theory: can we upgrade the minisuperspace approach to a systematic effective theory of quantum cosmology? 

\subsection*{Acknowledgements} 
We are thankful to Juan Maldacena, Alessandro Podo, Richard Schwartz, and Simone Speziale for useful discussions. 
The work of AN is partially supported by the US DOE (award number DE-SC011941) and by the Simons Foundation (award number 658906).

\appendix
\section{Symmetries of the 2+1 action}\label{app-a}

The gravitational action~\eqref{action-theta} is invariant under $SL(2,\mathbb{R})$. By exponentiating the three generators $\ell_2$, $\ell_3$, and $\ell_4$, we find that 
$\exp(\lambda \ell_2)$, $\exp(\epsilon \ell_3)$ and $\exp(\eta \ell_4)$ produce the coordinate transformations $\xi \rightarrow \xi'$, $\theta \rightarrow \theta'$, where, respectively, 
\begin{align} \label{first-sym}
\ell_2: \qquad & \xi ' \ = \ \xi \\
& \theta' \ = \ \theta + {\lambda}\\
\ell_3: \qquad &\cosh  \xi' \ = \ \cosh  \xi \cosh 2 \epsilon + \sinh  \xi \sinh 2 \epsilon \cos \theta\\
& \sin  \theta ' \ = \ \sin  \theta \frac{\sinh  \xi}{\sinh  \xi'} \\
\ell_4: \qquad &\cosh  \xi' \ = \ \cosh  \xi \cosh 2 \eta + \sinh  \xi \sinh 2 \eta \sin \theta \label{5-sym}\\
& \cos  \theta ' \ = \ \cos  \theta \frac{\sinh  \xi}{\sinh  \xi'}\, . \label{last-sym}
\end{align}

Now we want to show that, if the matter action does not depend on $\xi$ and $\theta$ and thus respects all the $SL(2,\mathbb{R})$ symmetries, all rotating solutions---i.e., those with a nonzero $\dot \theta$---are equivalent to a non-rotating solution. 
Let us consider a non-rotating (``nr") solution first, with $\theta = 0$. By varying~\eqref{action-theta} we get that the eom for $\xi$ simplifies to $(A^2 \dot \xi)\dot\ = 0$. So
\begin{equation} \label{soluu}
\theta_{\rm nr} = 0, \qquad \dot\xi_{\rm nr} = \frac{b}{A^2},
\end{equation}
is a solution of the system, for some constant $b$. The above non-rotating solution expresses the well-known result that, for a perfect fluid, the energy density associated with anisotropies $\sim \dot \xi^2$ scales as $A^{-2 d}$ in $d +1$ dimensions. We want to show that all other solutions with $\theta(t) \neq 0$ can be obtained by transforming the above. More generally, the eom for $\theta$ gives
\begin{equation}  \label{anysolution}
\dot \theta = \frac{J}{A^2 (\sinh  \xi)^2},
\end{equation}
for some constant $J$, which represents the conserved charge associated with rotations, that is, the angular momentum.

Now, let us apply the transformation $l_4$ to the solution~\eqref{soluu} with some parameter $\eta$. By using~\eqref{5-sym} and~\eqref{last-sym} we obtain a new solution $\xi(t)$, $\theta(t)$ with
\begin{align}
\cosh  \xi(t) \ =& \ \cosh  \xi_{\rm nr}(t) \cosh 2 \eta \label{second2} \\
\theta(t) \ =& \ \arccos\left(\frac{\sinh \xi_{\rm nr}(t)}{\sinh \xi(t)}\right)
\end{align}
Differentiating the second expression with respect to time and using~\eqref{second2} we find 
\begin{equation}
\dot \theta  = \frac{ \dot \xi_{\rm nr} \sinh 2 \eta }{1 - (\cosh \xi_{\rm nr})^2 (\cosh 2 \eta)^2}  = \frac{b  \sinh 2 \eta}{A^2 (\sinh \xi)^2}\, ,
\end{equation}
where~\eqref{soluu} and~\eqref{second2} have been used in the last step. It is clear that, by choosing $\eta$ such that $b  \sinh 2 \eta = J$,  any solution of the type~\eqref{anysolution} can be reproduced. One can then combine this with a constant rotation to fix the correct initial condition for $\theta(t)$. 

\section{Setting charges to zero by symmetry transformations} \label{charges}

As we mentioned in the main text, there are general arguments that indicate that, given a continuous symmetry group $G$ and its associated conservation laws, for generic initial conditions one can perform symmetry transformations that set to zero all charges apart from those in the Cartan subalgebra---the subalgebra that generates the maximal abelian subgroup of $G$ 
\footnote{This is, in fact, an upper bound on the number of nonzero charges that can survive. If the system's degrees of freedom make up a particularly small representation of $G$, one might be able to set even more charges to zero. For example, for a single particle in a central potential in $d$ dimensions, the phase space is parametrized just by two $d$-vectors $\vec q$ and $\vec p$, and, up to rotations, for generic initial conditions there is only {\em one} nonzero angular momentum---that which acts in the $\vec q$-$\vec p$ plane.}. 

In our specific cases, $G= SL(d,\mathbb{R})$  or $SO(d)$ acting on $h_{ij}(t)$ as
\be
h \to g^T \cdot h \cdot g \; , \label{transform h}
\ee
we can verify this claim explicitly.
Let us rewrite the above transformation law at the infinitesimal level using $g=\exp(\alpha^a T_a) \simeq 1 +\alpha^a T_a $, where the $T$'s are the generators of $G$ and the $\alpha$'s are transformation parameters. We get
\be
\delta h = \alpha^a (h \cdot T_a + T^T _a \cdot h) \; .
\ee

It is particularly convenient to work in the Hamiltonian formalism, where all symmetries $g \in G$ correspond to canonical transformations under which the Hamiltonian is invariant. In particular, for $g$ to be a canonical transformation to begin with, it must act on the conjugate momenta $p^{ij}$ as
\be
p \to  g^{-1} \cdot p \cdot \big( g^T \big)^{-1} \; ,
\ee 
so that
\be
\delta p = - \alpha^a (p \cdot T^T_a + T_a \cdot p) \; .
\ee

Now, in phase-space the above infinitesimal canonical transformation must be generated by a phase-space function $Q = Q(h, p) = \alpha^a \tau_a(h,p)$ as
\be
\delta h = \alpha^a \{ h , \tau_a(h,p) \} \; , \qquad \delta p = \alpha^a \{ p , \tau_a(h,p) \}  \; ,
\ee
where $\{\cdot , \cdot \}$ are the Poisson brackets. The $\tau_a$'s are the conserved charges associated with the generators $T_a$, expressed as  functions of $h_{ij}$ and $p^{ij}$. After trial and error, it is easy to convince oneself that they must take the form
\be
\tau_a(h,p) = 2 \, {\rm tr} \big( h \cdot T_a \cdot p \big) \; .
\ee  

Now the question is: assuming we start with generic initial conditions $h_{ij}(0)$, $p^{ij}(0)$, how many $\tau_a(h,p)$'s can we set to zero using only symmetry transformations?

There are two distinct cases:
\begin{itemize}
\item
$G = SL(d,\mathbb{R})$: In this case we can perform a transformation that sets $h_{ij}(0)$ proportional to the identity matrix. After we do so, we have
\be
\tau_a(h,p) \propto  {\rm tr} \big(  T_a \cdot p (0) \big) \; ,
\ee
and we are still allowed to perform $SO(d)$ rotations, since those are---by definition---the transformations in \eqref{transform h} that don't change the identity matrix. With these, we can make $p^{ij}(0)$ diagonal,
\be
p(0) = {\rm diag}(P^1, \dots, P^d)
\ee
We thus get
\be
\tau_a(h,p) \propto \sum_i (T_a)^i {}_{i} P^i  \; ,
\ee
and so, according to the classification of Sec.~\ref{symmetries}, we have that the only nonzero charges are those associated with the diagonal shears, which, indeed,  make up the Cartan subalgebra of $SL(d,\mathbb{R})$.

\item
$G = SO(d)$: In this case the $T_a$ generators are anti-symmetric and, using cyclicity of the trace and the fact that $h$ and $p$ are symmetric, we get
\be \label{tau rotations}
\tau_a(h,p) = {\rm tr} \big( T_a \cdot [\, p,h \, ] \big) \; .
\ee
Now, the $[\, p,h \, ]$ commutator is an anti-symmetric matrix, and can thus be block-diagonalized (at $t=0$) by a suitable rotation in $SO(d)$. The resulting block-diagonal (anti-symmetric) matrix has $\lfloor d/2 \rfloor$ antisymmetric $2 \times 2$ blocks  along the diagonal, and zeroes everywhere else. Plugging this structure into \eqref{tau rotations}
and parametrizing the rotation generators as in \eqref{parametrize rotations}, we see that the only nonzero charges that survive are those associated with the rotations that act within each of the aforementioned $2\times2$ blocks. These are precisely the generators of the Cartan subalgebra of $SO(d)$.

\end{itemize}

\section{Rotating cosmologies in 3+1 dimensions} \label{3D rotation}
\subsection{Parameterization of the metric}
We still write the spatial metric, in matrix form, as 
\begin{equation} 
h(t)= R^T(t) \cdot D(t) \cdot R (t)\, ,
\end{equation}
where $R$ is a rotation and $D$ a diagonal matrix. We choose to write the rotation as
\begin{equation}
R (t)= e^{i\,  \vec \theta (t)\cdot \vec J}\, ,
\end{equation}
where $(J_i)_{jk} = i \epsilon_{ijk}$ and $[J_i, J_j] = i \epsilon_{ijk} J_k$. We find that the time derivative of $R$ can be written as 
\begin{equation}
\dot R = i R P_+ = i P_- R\, ,
\end{equation}
where 
\begin{align}
P_\pm &= \vec E_\pm \cdot \vj \\
&\equiv  \dot \vt \cdot \vec J + \frac{\sin |\vt | - |\vt |}{|\vt |}\Bigg[ \dot \vt \cdot \vec J - \frac{\big(\vt \cdot \dot \vt \big)\big(\vt \cdot \vj \, \big)}{|\vt |^2}\Bigg]  \pm \Bigg[\frac{1 - \cos |\vt |}{|\vt|^2}\Bigg] \Big(\vt \wedge \dot \vt\Big) \cdot \vj
\end{align}

In the gravitational action 
\begin{equation} \label{bianchiaction2}
    S_{\rm grav}= \frac{1}{16 \pi G}\int  \frac{dt }{4 N} \sqrt{h } \left(h^{il} h^{jm} - h^{ij} h^{lm}\right) \dot{h}_{ij} \dot{h}_{lm} \; ,
\end{equation}
there are two different structure. For that involving  $h^{ij} \dot h_{ij}$,  only the diagonal part survives,
\begin{equation}
 h^{ij} \dot h_{ij} = {\rm Tr}\left(D^{-1} \dot D\right)\, .
 \end{equation}
 The other term is more complicated. We find
 \begin{equation}
h^{ij} \dot h_{jk}h^{kl} \dot h_{li} = -2 {\rm Tr}\left( P_-^2 \right) + 2 {\rm Tr}\left( D^{-1} P_- D P_- \right) + {\rm Tr}\left( D^{-1} \dot D D^{-1} \dot D\right).
\end{equation}
The first term represents the kinetic terms and the self interactions of the $\theta$-fields. Because ${\rm Tr}(J_l J_m)= 2 \delta_{lm}$, we get 
\begin{equation}
 {\rm Tr}\left( P_-^2 \right) = 2 \vec E_- \cdot \vec E_-
 \end{equation}
The second term represents the coupling between the shear terms on the diagonal matrix and the angles. In order to calculate this we introduce the matrices
\begin{equation}
M_1 = \begin{pmatrix} 1&0&0\\ 0&0&0 \\0&0&0 \end{pmatrix}\, , \quad M_2 = \begin{pmatrix} 0&0&0\\ 0&1&0 \\0&0&0 \end{pmatrix}\, , \quad M_3 = \begin{pmatrix} 0&0&0\\ 0&0&0 \\0&0&1 \end{pmatrix}\, .
\end{equation}
So that we can write the diagonal piece as
\begin{equation}
D(t) = \alpha_i(t) M_i\, .
\end{equation}
The coupling term above contains $M$ and $J$ matrices alternated inside the trace. We find
\begin{equation}
{\rm Tr}\left(J_i M_j J_k M_l\right) = \delta_{ik} |\epsilon_{i j l}|
\end{equation}
where no summation over $i$ is intended. We get
\begin{align}
 {\rm Tr}\left( D^{-1} P_- D P_- \right)& = E_1^2\left(\alpha_2 \alpha_3^{-1} + \alpha_3 \alpha_2^{-1}\right) + 
  E_2^2\left(\alpha_3 \alpha_1^{-1} + \alpha_1 \alpha_3^{-1}\right) +E_3^2\left(\alpha_1 \alpha_2^{-1} + \alpha_2 \alpha_1^{-1}\right) \, .
\end{align}
Finally the gravitational action reads
\begin{align}
 S= &\frac{1}{16 \pi G}\int  \frac{dt }{2 N}\big[- \dot \alpha_1 \dot \alpha_2 \alpha_3 -  \dot \alpha_1  \alpha_2 \dot \alpha_3- \alpha_1 \dot \alpha_2 \dot \alpha_3 - 2 \alpha_1 \alpha_2 \alpha_3 E_-^2  \\ 
  &+E_1^2 \alpha_1(\alpha_2^2 + \alpha_3^2) +
 E_2^2 \alpha_2(\alpha_1^2 + \alpha_3^2) + E_3^2 \alpha_3(\alpha_1^2 + \alpha_2^2) \big ]\, .
 \end{align}

\section{More on the solid's Lagrangian and Hamiltonian} \label{app_3}

As we discussed, in order to build the Lagrangian of a solid it is useful to introduce the Lorentz-scalar matrix $B^{IJ} \equiv \partial_{\mu} \phi^I \partial^{\mu} \phi^J $.  In $d$ spatial dimensions $ I, J ...$ take values between $1$ and $d$ and the Lagrangian is a general function of the invariants built with $B^{IJ}$, 
\begin{align} \label{eq:Action_solid}
    S_{\rm solid}= -\int d^d x \, dt \, N \sqrt{h}\ F\left([B], [B^2], \dots, [B^d]\right),  
\end{align}
where $[ \cdots]$ is shorthand for the trace. 

By using the ADM form for the metric~\eqref{adm}, we can perfor a useful $d+1$ decomposition of such a quantity:
\begin{align}
    B^{IJ} &=- \frac{1}{N^2} \left( \dot{\phi}^I - N^i \partial_i \phi^I \right)\left( \dot{\phi}^J - N^j \partial_j \phi^J \right) + h^{ij}\partial_i \phi^I\partial_j \phi^J \\[2mm]
           &\equiv - V^I V^J+ b^{IJ}\, .
\end{align}
Note that on the second line we have implicitly defined $V^I = N^{-1}( \dot{\phi}^I - N^i \partial_i \phi^I)$ and $b^{IJ} = h^{ij}\partial_i \phi^I\partial_j \phi^J $.

The perfect fluid limit of~\eqref{eq:Action_solid} corresponds to $F$ depending only on the determinant of $B^{IJ}$, 
\begin{equation} 
 \label{eq:Action_fluid}
    S_{\rm fluid}= -\int d^d x \, dt \, N \sqrt{h}\ F\left(\det B^{IJ}\right)\, . 
\end{equation}
Such a determinant can be expressed as
\begin{align}
    \det B^{IJ} &= \frac{1}{d\,!} \, \epsilon_{I_1 I_2 \dots I_d} \, \epsilon_{J_i J_2\dots J_d} \left(- V^{I_1} V^{J_1}+ b^{I_1 J_1}\right) \cdots 
      \left(- V^{I_d} V^{J_d}+ b^{I_d J_d} \right) \\[2mm]
    &=  \frac{1}{d\,!} \, \epsilon_{I_1 I_2 \dots I_d} \, \epsilon_{J_i J_2\dots J_d} \left(- d \, V^{I_1} V^{J_1} b^{I_2 J_2}\cdots b^{I_d J_d}
    + b^{I_1 J_1}\cdots b^{I_d J_d} \right)\\[2mm]
       &= (\det b^{IJ}) \left[1 - (b^{-1})_{IJ} V^I V^J \right]\, .
\end{align}

In order to write the Hamiltonian of the solid one should calculate the conjugate momentum $\pi_I = \partial {\cal L}/\partial \dot \phi_I$ and invert this relation in favor of $\dot \phi_I$. Such explicit calculation does not seem to be possible except in sporadic cases. One notable case is that of \emph{non-relativistic fluid}, i.e., when $F = (\det B^{IJ})^{1/2}$. Then we get
\begin{align} \label{conj-mom}
    \pi_I  = 2\sqrt{h}  \, \det b \, F'(\det B^{IJ})  (b^{-1})_{IJ} V^J 
\end{align}
By using the ansatz $V^I = A\, b^{IJ} \pi_J$ with $A$ to be determined, the above relation can be inverted to give 
\begin{align}
    V^I=\frac{b^{IJ} \pi_J}{\sqrt{ h\det b + b^{KL} \pi_K \pi_L}}\, ,
\end{align}
i.e.
\begin{align} \label{eq_inversion}
    \dot{\phi}^I = N \frac{b^{IJ} \pi_J}{\sqrt{ h\det b + b^{KL} \pi_K \pi_L}} + N^i \partial_i \phi^I \, .
\end{align}
Now we can express the Hamiltonian of the non-relativistic fluid, 
\begin{align}    \label{eq:Hamilt_fluid}
    H_{\rm fluid} &= \int d^d x\left(N {\cal H}^{\rm fluid} + N^i {\cal H}_i^{\rm fluid}\right),
      \end{align}
      with
\begin{align}
 {\cal H}^{\rm fluid}   & = \sqrt{h\det b + b^{KL} \pi_K \pi_L} \, ,\\[2mm]
  {\cal H}_i^{\rm fluid} & = \pi_I \partial_i \phi^I\, .  
\end{align}

\subsection{Momentum constraint}
It is useful to look at the momentum constrain for the solid. In order to obtain its explicit form in the Hamiltonian formalism we should be able to invert for $\dot \phi_I$ as in~\eqref{eq_inversion}, which does not seem to be doable in general. We can however calculate the momentum constraint in the Lagrangian formalism, by simply varying the action~\eqref{eq:Action_solid} with respect to $N^i$. 

First we note that 
\begin{equation}
\frac{\partial \left[ B^n\right]}{\partial N^i} = \frac{n}{N} \left(B^{n-1}\right)^{IJ}\left(\partial_i \phi^I V^J + \partial_i \phi^J V^I\right)\, ,
\end{equation}
where $\left(B^{n-1}\right)^{IJ}$ is the $(n-1)th$ power of the matrix $B^{IJ}$ (without tracing). This gives
\begin{equation}
\frac{\partial{\cal L}}{\partial N^i} = \sum_{n = 1}^d n \sqrt{h} \, F_{[B^n]}  \left[B^{n-1}\right]^{IJ}\left(\partial_i \phi^I V^J + \partial_i \phi^J V^I\right)=0\, .
\end{equation}
A simple and general solution of the above equation  is $V^I = 0$.  That is, 
\begin{equation} 
\dot{\phi}^I - N^i \partial_i \phi^I = 0\, .
\end{equation}
In unitary gauge, $x^i = \phi^I$, this simply reduces to the condition $N^i=0$.

We thus see that choosing the unitary gauge  always implies $N^i=0$ as a consequence of the momentum constraint, as anticipated in the main text.

 \section{Laplace-Beltrami operator in mini-superspace} \label{quantum}

We consider a general action of the type
\be \label{free particle 2}
S  \propto \frac12 \int \frac{dt }{N} \, G_{AB} (X) \dot X^{A} \dot X^B \; ,
\ee
and aim to calculate the Laplace-Beltrami operator associated with the metric $G_{AB}$, 
 \begin{equation} \label{LB}
  \nabla^2 = \frac{1}{\sqrt{-\det G_{AB}}} \partial_A \ \sqrt{-\det G_{AB}} \ G^{AB} \partial_B \, .
\end{equation}
In the absence of matter fields and up to an irrelevant normalization factor, the metric $G_{AB}$ can be read off eq.~\eqref{bianchiaction}, 
\begin{equation} \label{gdown}
G_{AB} =  h^{1/2}  \left(h^{ik} h^{jl} - h^{ij} h^{kl}\right),  
\end{equation}
where $h = \det h_{ij}$ and (lower) capital latin letters $A$, $B$ stand for pairs of (upper) symmetric indices, $(ij)$ and $(kl)$ respectively. The switch between lower and upper indexes is due to the fact that we are considering the (covariant) spatial metric as a generalized coordinate in field space, $h_{ij} \sim X^A$. 

A simple scaling argument shows that 
\begin{equation}
- \det G_{AB} = h^{-3/2}\, ,
\end{equation}
up to a positive constant factor, which cancels in~\eqref{LB}. The inverse of $G_{AB}$, on the other hand, is given in $d+1$ spacetime dimensions by
\begin{equation} \label{gup}
G^{BC} = \frac{1}{2 \, h^{1/2}}\left(h_{km} h_{ln} + h_{kn} h_{ml} - \frac{2}{d-1} \, h_{kl} h_{mn} \right),
\end{equation} 
where, $B=(kl)$ and $C=(mn)$. One can check that by multiplying the RHS of~\eqref{gdown} by that of~\eqref{gup} one obtains the identity matrix $\delta_A^C$, which, expressed in spatial $i,j, \dots$ indices,  is just the symmetrized product of Kronecker delta's, 
\begin{equation}
\frac12  \left(h^{ik} h^{jl} - h^{ij} h^{kl}\right)\ \left(h_{km} h_{ln} + h_{kn} h_{ml}- \frac{2}{d-1}\,  h_{kl} h_{mn} \right) \ = \
 \frac{1}{2} \left( \delta^i_m\delta^j_n + \delta^i_n\delta^j_m \right)\, .
\end{equation}

There are now all the ingredients to calculate~\eqref{LB},
\begin{align}
    \nabla^2 &=  h^{3/4} \frac{\partial}{\partial h_{ij}} \ h^{-5/4}\left(h_{ik} h_{jl} - \frac{1}{d-1}\,  h_{ij}h_{kl}\right)\\
            &= h^{-1/2} \left[ \left( h_{ik}h_{jl} - \frac{1}{d-1}\,  h_{ij}h_{kl} \right)\frac{\partial}{\partial h_{ij}}  \frac{\partial}{\partial h_{kl}} + \frac{2d^2 - 2d - 3}{4(d-1)}\,  h_{ij} \frac{\partial}{\partial h_{ij}}\right] 
\end{align}

For $d=2$ this expression gives the pure-gravity Laplacian displayed in eq.~\eqref{PG}. Alternatively, still in $d=2$, we can parametrize $h_{ij}$ in terms of the $A$, $\xi$, and $\theta$ variables and obtain 
\begin{align} \label{PG2}
    \nabla^2 = \frac{1}{2 A^2 } \left[    -\frac{A^2}{4} \partial_A^2    -\frac{A}{2} \partial_A +\frac{1}{ \tanh \xi} \partial_{\xi}   + \partial_{\xi}^2    +\frac{1}{\sinh^2 \! \xi} \partial_{\theta}^2  \right] \, .
\end{align}

As commented in Sec.~\ref{quantum2}, the structure of the above operators changes when including additional matter degrees of freedom in the theory. This is why the pure-gravity Laplacian~\eqref{PG2} is different than that in the presence of a scalar field~\eqref{not-so-mini}.

\bibliographystyle{utphys}
\bibliography{biblio.bib}
\end{document}